\documentclass[prb,superscriptaddress,twocolumn,letterpaper]{revtex4-1}
\usepackage{bm,graphicx,graphics,amsmath,amssymb,bm,epsfig,color}
\usepackage{euscript,tabularx}
\usepackage{longtable}
\usepackage[utf8]{inputenc}
\usepackage{MnSymbol,wasysym,stmaryrd,ifsym,rotating,pifont}

\def \pka {\text{\ding{192}}}
\def \pkb {\text{\ding{193}}}
\def \pkc {\text{\ding{194}}}
\def \pkd {\text{\ding{195}}}

\def \pkA {\text{\ding{202}}}
\def \pkB {\text{\ding{203}}}
\def \pkC {\text{\ding{204}}}
\def \pkD {\text{\ding{205}}}

\def\mathbi#1{\textbf{\em #1}}
\def\<{\left\langle}
\def\>{\right\rangle}
\def\op#1{{\widehat{\bm#1}}}
\def\unit#1{{\hat{\bm#1}}}
\def\cc#1{{\overline{\bm#1}}}
\def\dv{\!\!\!\!d^2\!\!\rho dz\,\,}

\begin{document}

\title{Identification and selection rules of the spin-wave eigen-modes
  in a normally magnetized nano-pillar}

\author{V.V. Naletov} \affiliation{Service de Physique de l'\'Etat
  Condens\'e (CNRS URA 2464), CEA Saclay, 91191 Gif-sur-Yvette,
  France} \affiliation{Physics Department, Kazan Federal University,
  Kazan 420008, Russian Federation}

\author{G. de Loubens} \email[Main co-author: ]{gregoire.deloubens@cea.fr}
\affiliation{Service de Physique de l'\'Etat Condens\'e (CNRS URA
  2464), CEA Saclay, 91191 Gif-sur-Yvette, France}

\author{G. Albuquerque} \affiliation{In Silicio, 730 rue Ren\'e Descartes
  13857 Aix En Provence, France}

\author{S. Borlenghi} \affiliation{Service de Physique de l'\'Etat
  Condens\'e (CNRS URA 2464), CEA Saclay, 91191 Gif-sur-Yvette,
  France}

\author{V. Cros} \affiliation{Unit\'e Mixte de Physique CNRS/Thales and
  Universit\'e Paris Sud 11, RD 128, 91767 Palaiseau, France}

\author{G. Faini} \affiliation{Laboratoire de Photonique et de
  Nanostructures, Route de Nozay 91460 Marcoussis, France}

\author{J. Grollier} \affiliation{Unit\'e Mixte de Physique CNRS/Thales
  and Universit\'e Paris Sud 11, RD 128, 91767 Palaiseau, France}

\author{H. Hurdequint} \affiliation{Laboratoire de Physique des
  Solides, Universit\'e Paris-Sud, 91405 Orsay, France}

\author{N. Locatelli} \affiliation{Unit\'e Mixte de Physique CNRS/Thales
  and Universit\'e Paris Sud 11, RD 128, 91767 Palaiseau, France}

\author{B. Pigeau} \affiliation{Service de Physique de l'\'Etat
  Condens\'e (CNRS URA 2464), CEA Saclay, 91191 Gif-sur-Yvette,
  France}

\author{A.~N. Slavin} \affiliation{Department of Physics, Oakland
  University, Michigan 48309, USA}

\author{V.~S. Tiberkevich} \affiliation{Department of Physics, Oakland
  University, Michigan 48309, USA}

\author{C. Ulysse} \affiliation{Laboratoire de Photonique et de
  Nanostructures, Route de Nozay 91460 Marcoussis, France}

\author{T. Valet} \affiliation{In Silicio, 730 rue Ren\'e Descartes
  13857 Aix En Provence, France}

\author{O. Klein} \email[Principal investigator and main co-author:
]{oklein@cea.fr} \affiliation{Service de Physique de l'\'Etat
  Condens\'e (CNRS URA 2464), CEA Saclay, 91191 Gif-sur-Yvette,
  France}

\date{\today}

\begin{abstract}
  We report on a spectroscopic study of the spin-wave eigen-modes
  inside an individual normally magnetized two layers circular
  nano-pillar (Permalloy$|$Copper$|$Permalloy) by means of a Magnetic
  Resonance Force Microscope (MRFM). We demonstrate that the observed
  spin-wave spectrum critically depends on the method of excitation.
  While the spatially uniform radio-frequency (RF) magnetic field
  excites only the axially symmetric modes having azimuthal index
  $\ell=0$, the RF current flowing through the nano-pillar, creating a
  circular RF Oersted field, excites only the modes having azimuthal
  index $\ell=+1$. Breaking the axial symmetry of the nano-pillar,
  either by tilting the bias magnetic field or by making the pillar
  shape elliptical, mixes different $\ell$-index symmetries, which can
  be excited simultaneously by the RF current. Experimental spectra
  are compared to theoretical prediction using both analytical and
  numerical calculations. An analysis of the influence of the static
  and dynamic dipolar coupling between the nano-pillar magnetic layers
  on the mode spectrum is performed.
\end{abstract}

\maketitle

\section{Introduction \label{sec:intro}}

Technological progress in the fabrication of hybrid nanostructures
using magnetic metals has allowed the emergence of a new science aimed
at utilizing spin dependent effects in the electronic transport
properties \cite{wolf01}. An elementary device of spintronics consists
of two magnetic layers separated by a normal layer. It exhibits the
well-known giant magneto-resistance (GMR) effect
\cite{baibich88,binasch89}, that is, its resistance depends on the
relative angle between the magnetic layers. Nowadays, this useful
property is extensively used in magnetic sensors \cite{dieny91,
  pannetier04}. The converse effect is that a direct current can
transfer spin angular momentum between two magnetic layers separated
by either a normal metal or a thin insulating layer
\cite{slonczewski96,berger96}. As a result, a spin polarized current
leads to a very efficient destabilization of the orientation of a
magnetic moment \cite{tsoi98}. Practical applications are the
possibility to control the digital information in magnetic random
access memories (MRAMs) \cite{albert00,grollier01} or to produce high
frequency signals in spin transfer nano-oscillators (STNOs)
\cite{kiselev03,rippard04}.

From an experimental point of view, the precise identification of the
spin-wave (SW) eigen-modes in hybrid magnetic nanostructures remains
to be done \cite{demidov04,woltersdorf07,loubens07,
  loubens07a,gubbiotti08,keatley08}. Of particular interest is the
exact nature of the modes excited by a current perpendicular-to-plane
in STNOs. Here, the identification of the associated symmetry behind
each mode is essential. It gives a fundamental insight about their
selection rules and about the mutual coupling mechanisms that might
exist intra or inter STNOs. It also determines the optimum strategy to
couple to the auto-oscillating mode observed when the spin transfer
torque compensates the damping, a vital knowledge to achieve phase
synchronization in arrays of nano-pillars \cite{slavin09}. These SW
modes also have a fundamental influence on the high frequency
properties of these devices and in particular on the noise of
magneto-resistive sensors \cite{nazarov02,stutzke03}.

A natural mean to probe SW modes in hybrid nanostructures is to use
their magneto-resistance properties. For instance, thermal SW can be
directly detected in the noise spectrum of tunneling
magneto-resistance (TMR) devices owing to their large TMR ratio
\cite{petit07,helmer10}. It is also possible to use spin torque driven
ferromagnetic resonance (ST-FMR)
\cite{tulapurkar05,sankey06,chen08,chen08a,biziere09,boone09,rippard10}.
In this approach, an RF current flowing through the magneto-resistive
device is used to excite the precession of magnetization and to detect
it through a rectification effect. Direct excitation of SW modes by
the RF field generated by micro-antennas and their detection through
dc rectification \cite{biziere08} or high-frequency GMR measurements
\cite{biziere08a} has also been reported in spin-valve sensors. In all
these experiments, the static magnetizations in the spin-valve have to
be misaligned in order for the magnetization precession to produce a
finite voltage. Because highly symmetric magnetization trajectories do
not produce any variation of resistance with time in some cases, a
third magnetic layer playing the role of an analyzer can be introduced
\cite{houssameddine07}. In ST-FMR, the non-collinearity of the
magnetizations is also required for the RF spin transfer excitation
not to vanish \cite{sankey06,chen08}. Moreover, the latter was never
directly compared to standard FMR, where a uniform RF magnetic field
is used to excite SW modes. Thus, although the voltage detection of SW
eigen-modes in hybrid nanostructures is elegant, one should keep in
mind that some of them might be hidden due to symmetry reasons.

Here, we propose an independent method of detecting the magnetic
resonance inside a spin-valve nanostructure. We shall use a Magnetic
Resonance Force Microscope (MRFM)
\cite{zhang96,wago98,jander01,charbois02a,klein08}. A first decisive
advantage of the MRFM technique is that the detection scheme does not
rely on the SW spatial symmetry because it measures the change in the
longitudinal component of the magnetization. Like a bolometric
detection, mechanical based FMR detects \emph{all} the excited SW
modes, independently of their phase \cite{loubens05,naletov07}. A
second decisive advantage is that MRFM is a very sensitive technique
that can measure the magnetization dynamics in nanostructures buried
under metallic electrodes \cite{loubens09,pigeau10,pigeau11}. Indeed,
the probe is a magnetic particle attached at the end of a soft
cantilever and is coupled to the sample through the dipolar
interaction.

In our roadmap to characterize the nature of the auto-oscillation
modes in STNOs, we report in this work on a comprehensive
identification of the SW eigen-modes in the simplest possible
geometry: the normally magnetized circular spin-valve nano-pillar.
This configuration is obtained by saturating the device with a large
external magnetic field oriented perpendicular to the layers. Thanks
to the preserved axial symmetry, a simplified spectroscopic signature
of the different SW eigen-modes is expected. This identification is
achieved experimentally from a comparative spectroscopic study of the
SW eigen-modes excited either by an RF current flowing perpendicularly
through the nano-pillar, as used in ST-FMR, or by a homogeneous RF
in-plane magnetic field, as used in conventional FMR. It shall be
developed as follows. In section II, we present the MRFM setup and the
experimental protocol used to perform SW spectroscopy in a
spin-valve. We show that the SW spectrum excited by a homogeneous RF
magnetic field is distinct from the SW spectrum excited by an RF
current flowing through the nano-pillar. In section III, we perform
unambiguous assignment of the resonance peaks to the different layers
by experimental means. We determine which layer contributes mostly to
the observed resonant signals by adding a direct current through the
nano-pillar, that produces opposite spin transfer torques on each
magnetic layer. In section IV, we analyze the spectra by theoretical
means using both a two-dimensional analytical formalism and a
three-dimensional micromagnetic simulation package, SpinFlow 3D. By
careful comparison of the measured spectra to the calculations, the
nature of the SW dynamics in the system is identified and the
selection rules for SW spectroscopy in perpendicularly magnetized
spin-valve nanostructures are established. This result is completed in
section V by a study of the influence of symmetry breaking on the
selection rules. This is obtained experimentally by introducing a tilt
angle of the applied magnetic field, and in simulations by changing
the shape of the nano-pillar. In the conclusion, we emphasize the
importance of this work for phase synchronization of STNOs. The paper
is arranged in such a fashion so as to present the main results in the
body of the text. A comprehensive appendix has been put at the end of
the paper, where the details of the introduced material are developed.

\section{Ferromagnetic resonance force spectroscopy \label{sec:MRFM}}

This section starts with a description of the nano-pillar sample,
followed by a description of the MRFM instrument used for this
spectroscopic study. Then, we compare the experimental SW spectra
excited by an RF current flowing perpendicularly through the
nano-pillar, as used in ST-FMR, and by a uniform RF magnetic field
applied parallel to the layers, as used in standard FMR.

\subsection{The lithographically patterned
  nanostructure \label{sec:litho}}

\begin{figure}
  \includegraphics[width=8.5cm]{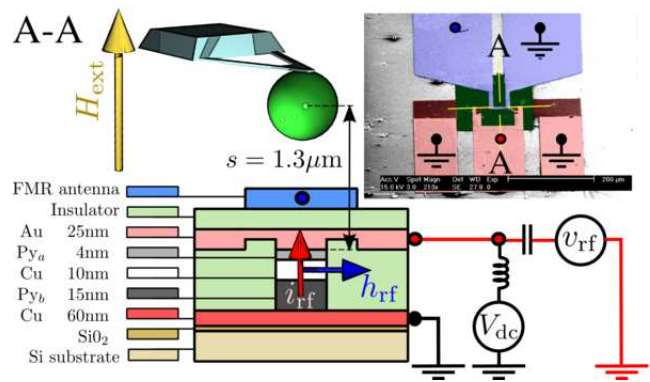}
  \caption{(Color online) Schematic representation of the experimental
    setup used for this comparative spin-wave spectroscopic study. The
    magnetic sample is a circular nano-pillar comprising a thin Py$_a$
    and a thick Py$_b$ magnetic layers separated by a Cu spacer. It is
    saturated by a large magnetic field $\mathbi{H}_{\rm ext}$ applied
    along its normal axis. A cantilever with a magnetic sphere
    attached at its tip monitors the magnetization dynamics inside the
    buried structure. The inset is a microscopy image (top view) of
    the two independent excitation circuits: in red the circuit
    allowing the injection of an RF current perpendicular-to-plane
    through the nano-pillar ($i_\text{rf}$, red arrow); in blue the
    circuit allowing the generation of an RF in-plane magnetic field
    ($h_\text{rf}$, blue arrow). The nano-pillar is at the center of
    the yellow cross-hair. The main figure is a section along the
    $A-A$ direction.}
  \label{fig:01}
\end{figure}

The spin-valve structure used in this study is a standard Permalloy
(Ni$_{80}$Fe$_{20}$=Py) bi-layer structure sandwiching a 10~nm copper
(Cu) spacer: the thicknesses of the thin Py$_a$ and the thick Py$_b$
layers are respectively $t_a=4$~nm and $t_b=15$~nm. Special care has
been put in the design of the microwave circuit around the
nano-pillar. The inset of FIG.~\ref{fig:01} shows a scanning electron
microscopy top view of this circuit. The nano-pillar is located at the
center of the cross-hair, in the middle of a highly symmetric pattern
designed to minimize cross-talk effects between both RF circuits shown
in blue and red, which provide two independent excitation means.

The nano-pillar is patterned by standard e-beam lithography and
ion-milling techniques from the extended film, (Cu60 $|$ Py$_b$15 $|$
Cu10 $|$ Py$_a$4 $|$ Au25) with thicknesses expressed in nm, to a
nano-pillar of nominal radius 100~nm. A precise control allows to stop
the etching process exactly at the bottom Cu layer, which is
subsequently used as the bottom contact electrode. A planarization
process of a polymerized resist by reactive ion etching enables to
uncover the top of the nano-pillar and to establish the top contact
electrode. The top and bottom contact electrodes are shown in red tone
in FIG.~\ref{fig:01}. These pads are impedance matched to allow for
high frequency characterization by injecting an RF current
$i_\text{rf}$ through the device. The bottom Cu electrode is grounded
and the top Au electrode is wire bounded to the central pin of a
microwave cable. Hereafter, spectra associated to SW excitations by
this part of the microwave circuit will be displayed in red tone. The
nano-pillar is also connected through a bias-T to a dc current source
and to a voltmeter through the same contact electrodes, which can be
used for standard current perpendicular to the plane (CPP-GMR)
transport measurements \cite{footnoteGMR}. In our circuit, a positive
current corresponds to a flow of electrons from the Py$_b$ thick layer
to the Py$_a$ thin layer and stabilizes the parallel configuration due
to the spin transfer effect \cite{slonczewski96,berger96}. The studies
presented below will be limited to a dc current up to the threshold
current for auto-oscillations in the thin layer.

The originality of our design is the addition of an independent top
microwave antenna, whose purpose is to produce an in-plane RF magnetic
field $h_\text{rf}$ at the nano-pillar location. In FIG.~\ref{fig:01}
this part of the microwave circuit is shown in blue tone. The
broadband strip-line antenna consists of a 300~nm thick Au layer
evaporated on top of a polymer layer that provides electrical
isolation from the rest of the structure. The width of the antenna
constriction situated above the nano-pillar is 10~$\mu$m. Injecting a
microwave current from a synthesizer inside the top antenna produces a
homogeneous in-plane linearly polarized microwave magnetic field,
oriented perpendicular to the stripe direction. Hereafter, spectra
associated to SW excitations by this part of the microwave circuit
will be displayed in blue tone.

\subsection{Mechanical-FMR \label{sec:mechanical}}

The nano-fabricated sample is then mounted inside a Magnetic Resonance
Force Microscope (MRFM), hereafter named mechanical-FMR
\cite{klein08}. The whole apparatus is placed inside a vacuum chamber
($10^{-6}$~mbar) operated at room temperature. The external magnetic
field produced by an electromagnet is oriented out-of-plane,
\textit{i.e.}, along the nano-pillar axis $\unit z$. The
mechanical-FMR setup allows for a precise control, within $0.2^\circ$,
of the polar angle between the applied field and $\unit z$. In our
study, the strength of the applied magnetic field shall exceed the
saturation field ($\approx 8$~kOe), so that the nano-pillar is studied
in the saturated regime.

The mechanical detector is an ultra-soft cantilever, an Olympus
Bio-Lever having a spring constant $k\approx5$~mN/m, with a 800~nm
diameter sphere of soft amorphous Fe (with 3\% Si) glued to its apex.
Standard piezo displacement techniques allow for positioning the
magnetic spherical probe precisely above the center of the
nano-pillar, so as to retain the axial symmetry. This is obtained when
the dipolar interaction between the sample and the probe is maximal,
by minimizing the cantilever resonance frequency, which is
continuously monitored \cite{loubens09}.

The mechanical sensor is insensitive to the rapid oscillations of the
transverse component in the sample, which occur at the Larmor
precession frequency, \textit{i.e.}, several orders of magnitude
faster than its mechanical resonances. The dipolar force on the
cantilever probe is thus proportional to the static component of the
magnetization inside the sample. For our normally magnetized sample,
this longitudinal component reduces to $M_z$.  We emphasize that for a
bi-layer system, the force signal integrates the contribution of both
layers. Moreover, the local $M_z(\bm{r})$ in the two magnetic layers
is weighted by the distance dependence of the dipolar coupling to the
center of the sphere. In our case though, where the separation between
the sphere and the sample is much larger than the sample dimensions,
one can neglect this weighting and the measured quantity simplifies to
the spatial average:
\begin{equation}\label{eq:average}
\langle {M}_z \rangle  \equiv \frac{1}{V}\int_V  {M}_z(\bm r) d^3\bm
r\, , 
\end{equation}
where the chevron brackets stand for the spatial average over the
volume of the magnetic body.

The mechanical-FMR spectroscopy presented below consists in recording
by optical means the vibration amplitude of the cantilever either as a
function of the out-of-plane magnetic field $H_\text{ext}$ at a fixed
microwave excitation frequency $f_\text{fix}$, or as a function of the
excitation frequency $f$ at a fixed magnetic field
$H_\text{fix}$. This type of spectroscopy is called cw, for continuous
wave, as it is monitoring the magnetization dynamics in the sample
under a forced regime. A source modulation is applied on the cw
excitation. It consists in a cyclic absorption sequence, where the
microwave power is switched on and off at the cantilever resonance
frequency, $f_c \approx 11.85$~kHz. The signal is thus proportional to
$\< \Delta {M}_z \> $, where $\Delta$ represents the difference from
the thermal equilibrium state. The source modulation enhances the
signal, recorded by a lock-in detection, by the quality factor
$Q\approx 2000$ of the mechanical oscillator. The force sensitivity of
our mechanical-FMR setup is better than 1~fN, corresponding to less
than $10^3$ Bohr magnetons in a bandwidth of one second
\cite{klein08}. We note that this modulation technique does not affect
the line shape in the linear regime, because the period of modulation
$1/f_c$ is very large compared to the relaxation times of the studied
ferromagnetic system \cite{klein03,klein04}. Moreover, we emphasize
that since the mechanical-FMR signal originates from the cyclic
diminution of the spatially averaged magnetization inside the whole
nano-pillar synchronous with the absorption of the microwave power, it
detects all possible SW modes without discrimination
\cite{loubens05,naletov07}.

Finally, we mention that the stray field produced by the magnetic
sphere attached on the cantilever does affect the detected SW
spectra. In our setup, the separation between the center of the
spherical probe and the nano-pillar is set to 1.3~$\mu$m (see
FIG.~\ref{fig:01}), which is a large distance considering the lateral
size of the sample. At such distance, the coupling between the sample
and the probe is weak \cite{klein08} as it does not affect the
profiles of the intrinsic SW modes in the sample. This is in contrast
with the strong coupling regime, where the stray field of the magnetic
probe can be used to localize SW modes below the MRFM tip
\cite{lee10}. For our mechanical SW spectrometer, the perturbation of
the magnetic sphere reduces to a uniform translation of all the peak
positions \cite{charbois02} by $-190$~Oe (see
section~\ref{sec:voltage}). In the following, all the SW spectra are
recorded with the magnetic sphere at the same exact position above the
nano-pillar.

\subsection{RF magnetic field vs. RF current excitations  \label{sec:hrfvsirf}}

\begin{figure}
  \includegraphics[width=8.5cm]{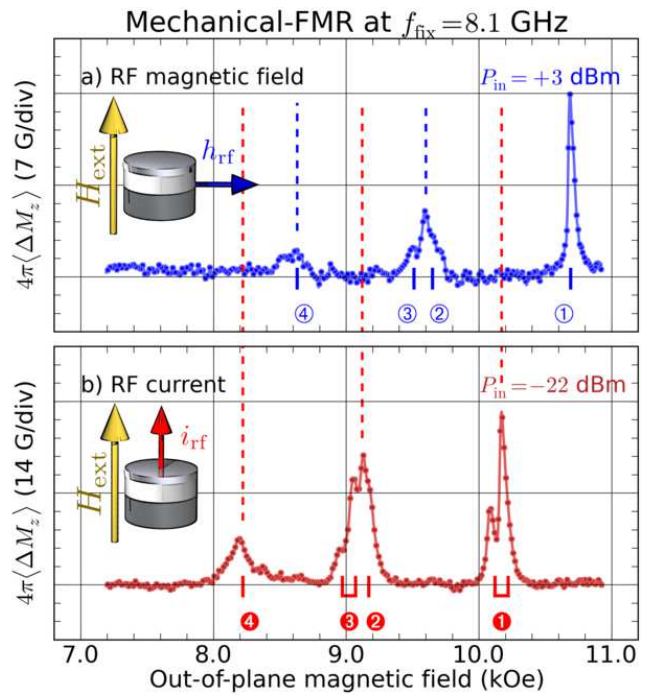}
  \caption{(Color online) Comparative spectroscopic study performed by
    mechanical-FMR at $f_\text{fix}=8.1$~GHz, demonstrating that
    distinct SW spectra are excited by a uniform in-plane RF magnetic
    field (a) and by an RF current flowing perpendicularly through the
    layers (b). The positions of the peaks are reported in
    Table~\ref{tab:eigen}.}
  \label{fig:02}
\end{figure}

The comparative spectroscopic study performed by mechanical-FMR at
$f_\text{fix}=8.1$~GHz on the normally magnetized spin-valve
nano-pillar is presented in FIG.~\ref{fig:02}. In these experiments,
there is no dc current flowing through the device, and the spectra are
obtained in the small excitation regime (precession angles less than
$5^\circ$, see appendix~\ref{app:Mz}). The upper panel (a) shows the
SW spectrum excited by a uniform RF magnetic field applied in the
plane of the layers, while the lower panel (b) displays the SW
spectrum excited by an RF current flowing perpendicularly through the
magnetic layers. The striking result is that these two spectra are
different: none of the SW modes excited by the homogeneous RF field is
present in the spectrum excited by the RF current flowing through the
nano-pillar, and vice versa.

Let us first focus on FIG.~\ref{fig:02}a, where the obtained
absorption spectrum corresponds to the so-called standard FMR
spectrum. Here, the output power of the microwave synthesizer at
$8.1$~GHz is set to $+3$~dBm, which corresponds to an amplitude of the
uniform linearly polarized RF magnetic field $h_\text{rf} \simeq
2.1$~Oe produced by the antenna (see appendix~\ref{app:Mz}). In this
standard FMR spectrum, only SW modes with non-vanishing spatial
average can couple to the homogeneous RF field excitation. In
field-sweep spectroscopy, the lowest energy mode occurs at the largest
magnetic field. So, the highest field peak at $H_{\pka}=10.69$~kOe
should be ascribed to the uniform mode. Since this peak is also the
largest of the spectrum, it corresponds to the precession of a large
volume in the nano-pillar, \textit{i.e.}, the thick layer must
dominate in the dynamics. In mechanical-FMR, a quantitative
measurement of the longitudinal magnetization is obtained
\cite{naletov03,loubens05} (see appendix~\ref{app:Mz}). The amplitude
of the peak at $H_{\pka}$ corresponds to $4 \pi\langle \Delta M_z
\rangle \simeq 14$~G, which represents a precession angle $\langle
\theta \rangle \simeq 3.1^\circ$. This sharp peak is followed by a
broader peak with at least two maxima at $H_{\pkb}=9.65$~kOe and
$H_{\pkc}=9.51$~kOe, and at lower field, by a smaller resonance around
$H_{\pkd}=8.64$~kOe. Among these other peaks, there is the uniform
mode dominated by the thin layer, which has to be identified and
distinguished from higher radial index SW modes.

Let us now turn to FIG.~\ref{fig:02}b, corresponding to the
spectroscopic response to an RF current of same frequency $8.1$~GHz
flowing perpendicularly through the nano-pillar. Here, the output
power of the microwave synthesizer is $-22$~dBm, which corresponds to
an rms amplitude of the RF current $i_\text{rf} \simeq 170$~$\mu$A
(see appendix~\ref{app:RF}). The SW spectrum is acquired under the
\emph{exact same conditions} as for standard FMR, \textit{i.e.}, the
spherical magnetic probe of the mechanical-FMR detection is kept at
the same location above the sample. The striking result is that the
position of the peaks in FIGS.~\ref{fig:02}a and \ref{fig:02}b do not
coincide. More precisely there seems to be a translational
correspondence between the two spectra, which are shifted in field by
about 0.5~kOe from each other. The lowest energy mode in the RF
current spectrum occurs at $H_{\pkA} =10.22$~kOe. This is again the
most intense peak, suggesting that the thick layer contributes to it,
and $ 4 \pi\langle \Delta M_z \rangle \simeq 26$~G, which represents a
precession angle $\langle \theta \rangle \simeq 4.2^\circ$. This main
resonance line is also split in two peaks, with a smaller resonance in
the low field wing of the main peak, about 100~Oe away. At lower
field, two distinct peaks appear at $H_{\pkB}=9.17$~kOe and
$H_{\pkC}=9.07$~kOe and another peak is visible at
$H_{\pkD}=8.22$~kOe.

The fact that the two spectra of FIGS.~\ref{fig:02}a and \ref{fig:02}b
are distinct implies that they have a different origin. It will be
shown in the theoretical section~\ref{sec:rules} that the RF field and
the RF current excitations probe two different azimuthal symmetries
$\ell$. Namely, only $\ell=0$ modes are excited by the uniform RF
magnetic field, whereas only $\ell=+1$ modes are excited by the
orthoradial RF Oersted field associated to the RF current
\cite{arias09}. The mutually exclusive nature of the responses to the
uniform and orthoradial symmetry excitations is a property of the
preserved axial symmetry, where the azimuthal index $\ell$ is a good
quantum number, \textit{i.e.}, different $\ell$-index modes are not
mixed and can be excited separately (see section~\ref{sec:normal}).

\section{Experimental analysis  \label{sec:exp}}

In this section, we first look at the effect of a continuous current
flowing through the nano-pillar on the SW spectra in order to
determine which layer contributes mostly to the resonant signals
observed in FIG.~\ref{fig:02}. Due to the asymmetry of the spin
transfer torque in each magnetic layer, the different SW modes are
influenced differently depending on the layer in which the precession
is the largest. Then, we briefly mention experiments, where
spectroscopy is performed by monitoring the dc voltage produced by the
magnetization precession in the hybrid nanostructure, and compared to
mechanical-FMR. Finally, the analysis of the frequency-field
dispersion relation and of the linewidth of the resonance peaks
enables to extract the gyromagnetic ratio and the damping parameters
in the thick and thin layers.

\subsection{Direct bias current  \label{sec:idc}}

\begin{figure}
  \includegraphics[width=8.5cm]{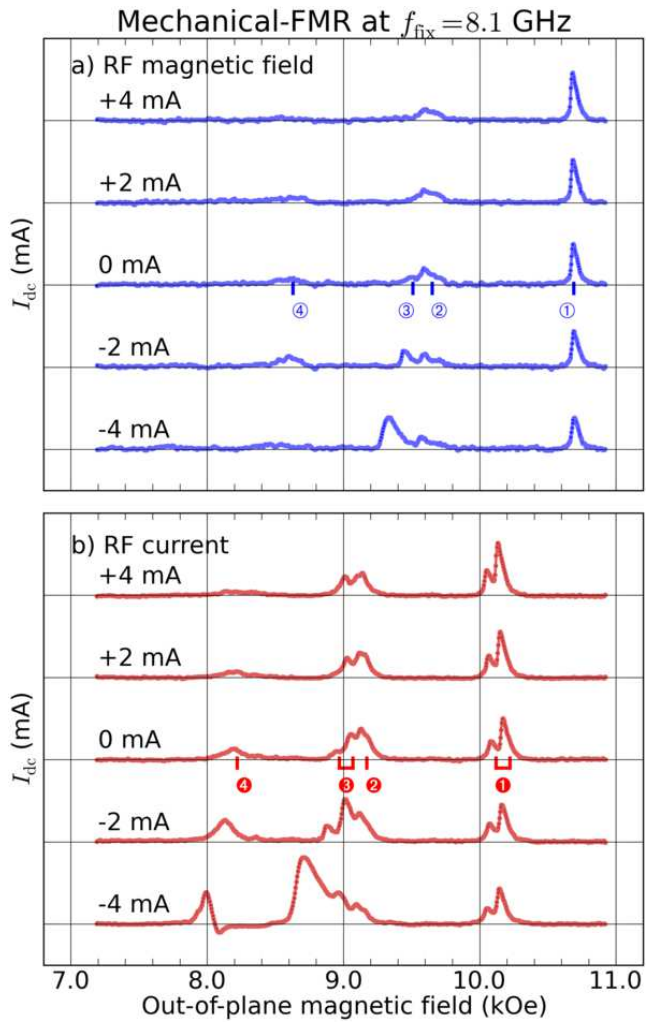}
  \caption{(Color online) Evolution of the SW spectra measured at
    $f_\text{fix}=8.1$~GHz by mechanical-FMR for different values of
    the continuous current $I_\text{dc}$ flowing through the
    nano-pillar. The panel (a) corresponds to excitation by a uniform
    RF magnetic field and the panel (b) to excitation by an RF current
    through the sample.}
  \label{fig:03}
\end{figure}

To gain further insight about the peak indexation, we have measured
the spectral evolution produced on the SW spectra of FIG.~\ref{fig:02}
when a finite dc current $I_\text{dc} \neq 0$ is injected in the
nano-pillar. We recall that for our sign convention, a positive dc
current stabilizes the thin layer and destabilizes the thick one due
to the spin transfer torque, and vice versa
\cite{slonczewski96,berger96}. The results obtained by mechanical-FMR
are reported in FIG.~\ref{fig:03}.

Let us first concentrate on FIG.~\ref{fig:03}a, in which the
excitation that probes the different SW modes is the same as in
FIG.~\ref{fig:02}a, \textit{i.e.}, a uniform RF magnetic field. Two
main features can be observed in the evolution of the SW spectra as
$I_\text{dc}$ is varied. First, the amplitude of the peak at
$H_{\pka}$ smoothly increases with the positive current and smoothly
decreases with the negative current. At the same time, the peak at
$H_{\pkc}$, which is about five times smaller than the peak at
$H_{\pka}$ when $I_\text{dc}=0$~mA, almost disappears for positive
current and strongly increases at negative current, until it becomes
larger than the other peaks when $I_\text{dc}=-4$~mA. These two
features are consistent with the effect of spin transfer if we ascribe
the peak at $H_{\pka}$ to the uniform mode of mostly the thick layer
and the peak at $H_{\pkc}$ to the one of mostly the thin layer. More
precisely, it is expected that in the sub-critical regime
($|I_\text{dc}|<I_\text{th}$, where $I_\text{th}$ is the threshold
current for auto-oscillations, $I_\text{th}<0$ for the thin layer and
$I_\text{th}>0$ for the thick layer), the damping scales as $\alpha
(1-I_\text{dc}/I_\text{th})$ \cite{sankey06,chen08} (see
appendix~\ref{app:motion}), where $\alpha$ is the Gilbert damping
parameter. It means that the linewidth of a resonance peak that is
favored by spin transfer should decrease as the current gets closer to
$I_\text{th}$, and that its amplitude, which scales as the inverse
linewidth, should increase.

Although the effect on the peak amplitude noted above is clear in
FIG.~\ref{fig:03}a, it is not on the linewidth. The reason is that in
this experiment, the strength of the driving RF magnetic field is kept
constant to $h_\text{rf} = 2.1$~Oe. As a result, the shape of the
growing peaks in FIG.~\ref{fig:03}a becomes more asymmetric, which is
a signature that the precession amplitude driven by the RF field is
strong enough to change the internal field by an amount of the order
of the linewidth. This leads to some foldover of the resonance line
\cite{anderson55,schlomann59a}, a non-linear effect for which details
are given in the appendix~\ref{app:Mz}. In other words, the distortion
of the line shape as the peak amplitude increases prevents to see the
diminution of its linewidth \cite{chen09}. It would be necessary to
decrease the excitation amplitude as the threshold current is
approached \cite{chen08} so as to maintain the peak amplitude in the
linear regime in order to reveal it.

The opposite signs of the spin transfer torques which influence the
dynamics in the thin and thick layers are thus clearly seen in
FIG.~\ref{fig:03}a. Their relative strengths can also be determined,
as the amplitude of the peak at $H_{\pkc}$ grows much faster with
negative current than the one of the peak at $H_{\pka}$ with positive
current. This is because the efficiency of the spin transfer torque is
inversely proportional to the thickness of the layer
\cite{slonczewski96,berger96}. Whereas the precession angle in the
thick layer does not vary much with $I_\text{dc}$ (from $\approx
2.5^\circ$ at $-4$~mA to $\approx 3.5^\circ$ at $+4$~mA), the
precession angle that can be deduced from $\langle \Delta M_z \rangle$
in the thin layer grows from almost zero at $I_\text{dc}=+4$~mA to
more than $6^\circ$ at $I_\text{dc}=-4$~mA.  Moreover, the peak
position $H_{\pkc}$ shifts clearly towards lower field as the negative
current is increased. This is due to the onset of spin transfer driven
auto-oscillations in the thin layer, which occurs at a threshold
current $I_\text{th}\lesssim -4$~mA and produces this non-linear shift
\cite{slavin09}. We note, that such a value for the threshold current
in the thin layer can be found from Slonczewski's model (see
appendix~\ref{app:motion}).

Let us now briefly discuss FIG.~\ref{fig:03}b, which shows the
dependence on $I_\text{dc}$ of the mechanical-FMR spectra excited by
an RF current excitation. A similar dependence on $I_\text{dc}$ of the
resonance peaks in translational correspondence with
FIG.~\ref{fig:03}a is observed. Again, a clear asymmetry is revealed
depending on the polarity of $I_\text{dc}$ and on the SW modes. The
double peak at $H_{\pkA}$ is favored by positive currents, hence it
should be ascribed to mostly the thick layer precessing, while the
double peak at $H_{\pkC}$ is strongly favored by negative currents,
hence it should be ascribed to mostly the thin layer
precessing. Moreover, a careful inspection shows that the peak
$H_{\pkB}$, which looks single at $I_\text{dc}=0$~mA, is actually at
least double. We will explain this splitting of higher harmonics modes
in section~\ref{sec:asym}.

To summarize, the passage of a dc current through the nano-pillar
enables to determine which layer mostly contributes to the observed SW
modes, owing to the asymmetry of the spin transfer effect.

\subsection{Voltage-FMR  \label{sec:voltage}}

Our experimental setup also allows to monitor the dc voltage produced
across the nano-pillar by the precession of the magnetization in the
bi-layer structure. A lock-in detection is used to measure the
difference of voltage across the nano-pillar when the RF is on and
off: $V_\text{dc}=V_\text{on}-V_\text{off}$. This can be done
\emph{simultaneously} to the acquisition of the mechanical-FMR signal,
in the exact same conditions (see FIG.~\ref{fig:01}). Since the
presentation of the experimental results requires a specific
discussion, the details as well as the graphs will be published
elsewhere. Here, we shall only reveal the three main features that can
be noticed in the voltage-FMR spectra.

First, even at $I_\text{dc}=0$, dc voltage peaks are produced across
the nano-pillar at the same positions as the mechanical-FMR peaks
observed in FIG.~\ref{fig:02}, with a difference of potential that
lies in the 10~nV range for the precession angles excited here. It is
ascribed to spin pumping and accumulation in the spin-valve hybrid
structure \cite{tserkovnyak05,costache06}. Second, these voltage
resonance peaks are signed, namely, the SW modes favored at
$I_\text{dc}<0$ in FIG.~\ref{fig:03}a (for which the thin layer is
dominating) produce a positive voltage peak, whereas those favored at
$I_\text{dc}>0$ (thick layer dominating) produce a negative voltage
peak. This difference between the thick and thin layer contributions
is ascribed to the asymmetry of the spin accumulation in the
multi-layer stack \cite{kupferschmidt06}. Third, the relative
amplitudes of the voltage-FMR peaks are different from the
mechanical-FMR ones. For instance, the voltage-FMR peak of the thin
layer at $H_{\pkc}$ is slightly \emph{larger} than the peak at
$H_{\pka}$ of the thick layer (and it has an opposite sign). This
illustrates an important difference between the two detection
schemes. While mechanical-FMR measures a quantity proportional to the
precessing volume, $\langle \Delta M_z \rangle$, the voltage-FMR
measures an interfacial effect. Therefore, when the same precession
angle is excited in both layers, the voltage-FMR signal associated to
each layer is approximately the same, whereas the mechanical-FMR
signal from the thin layer is roughly four times smaller than the one
from the thick layer, due to their relative thicknesses.

Finally, we mention that voltage-FMR spectroscopy can also record the
intrinsic FMR spectrum of the nano-pillar, \textit{i.e.}, in the
absence of the spherical MRFM probe above it. This enables to check
that the only effect introduced by the probe in mechanical-FMR is an
overall shift of the SW modes spectra to lower field without any other
distortion, and to quantify this shift, found to be $-190$~Oe
\cite{footnotestrayfield}.

\subsection{Gyromagnetic ratio \label{sec:gyro}}

A precise orientation of the applied magnetic field $\mathbi{H}_{\rm
  ext}$ along the normal $\unit z$ of the sample (polar angle
$\theta_H=(\unit z,\bm{H}_{\rm ext})=0$) enables a direct
determination of the modulus $\gamma$ of the gyromagnetic ratio
\cite{klein08}. By following the frequency-field dispersion relation
of the resonance peaks at $H_{\pka}$ and at $H_{\pkc}$ (from 4.5~GHz
to 8.1~GHz and from 6.2~GHz to 11~GHz, respectively) in our
nano-pillar, it is found that $\gamma=1.87\times
10^{7}$~rad.s$^{-1}$.G$^{-1}$ is identical in the thick and thin
layers. Moreover, the value of $\gamma$ measured in the nano-pillar is
the same as in the extended reference film (see
appendix~\ref{app:cavity} and Table~\ref{tab:param}), confirming that
the applied field is sufficient to saturate the two magnetic layers
and is precisely oriented along $\unit z$.

\begin{figure}
  \includegraphics[width=8.5cm]{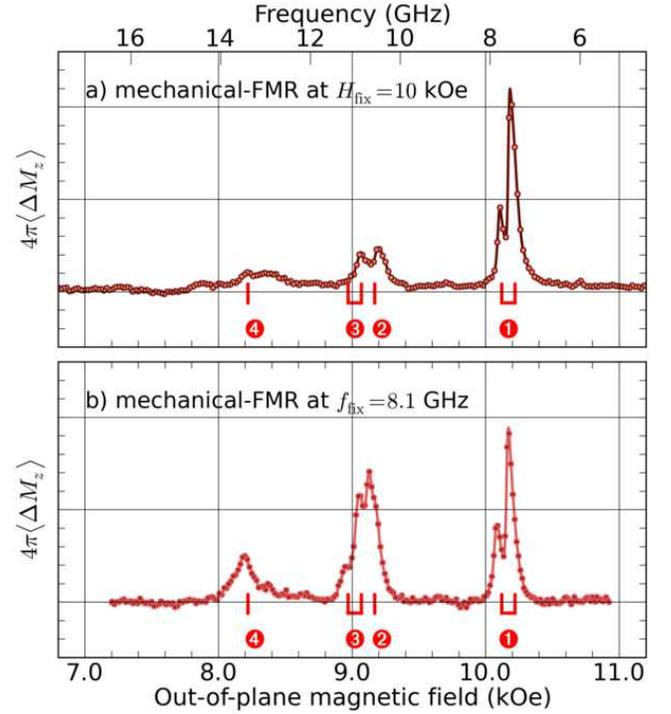}
  \caption{(Color online) Frequency-field dispersion relation: the top
    spectrum (a) is measured at fixed bias field $H_\text{fix}=10$~kOe
    by sweeping the frequency of the RF current $i_\text{rf}$ through
    the nano-pillar. The bottom spectrum (b) is the same as in
    FIG.~\ref{fig:02}b, and is obtained by sweeping the magnetic field
    at fixed frequency $f_\text{fix}=8.1$~GHz of $i_\text{rf}$. The
    top and bottom scales are in correspondence through the affine
    transformation $H_\text{ext}- H_\text{fix}=2\pi
    (f-f_\text{fix})/\gamma$.}
  \label{fig:04}
\end{figure}

The same result is obtained by following the evolution of the
frequency-field dispersion relation presented in
FIG.~\ref{fig:04}. Here, we take advantage of the broadband design of
the electrodes which connect the nano-pillar to measure the FMR
spectrum at fixed bias magnetic field, $H_\text{fix}=10$~kOe, by
sweeping the frequency of the RF current through it. The data are
plotted according to the frequency scale above FIG.~\ref{fig:04}a. At
constant magnetic configuration (above the saturation field,
\textit{i.e.}, $\gtrsim 8$~kOe), this frequency scale is in
correspondence with field-sweep experiments performed at fixed RF
frequency $f_\text{fix}=8.1$~GHz through the affine transformation
$H_\text{ext}- H_\text{fix}=2\pi (f-f_\text{fix})/\gamma$, as seen
from the field scale below FIG.~\ref{fig:04}b. This is a direct
experimental check of the equivalence between frequency and field
sweep experiments in the normally saturated state.

\subsection{Damping parameters \label{sec:damping}}

From the FMR data presented above, we can also directly extract the
damping parameters in each Permalloy layer. Indeed, in field-sweep
spectroscopy in the normal orientation ($\theta_H=0$), the full width
at half-maximum (FWHM) $\Delta H$ of a resonance line is proportional
to the excitation frequency $\omega/(2\pi)$ through the Gilbert
constant $\alpha$: $\Delta H=2\alpha(\omega/\gamma)$ (see
appendix~\ref{app:motion}).

The linewidth of the peak at $H_{\pka}$ associated to mainly the thick
layer in FIG.~\ref{fig:02}a is equal to $\Delta H_{\pka}=48$~Oe, which
corresponds to a damping $\alpha_{\pka}=0.88 \times 10^{-2}$. From the
same mechanical-FMR spectrum, the linewidth of the peak at $H_{\pkc}$,
associated to mainly the thin layer, cannot be easily extracted due to
the proximity of the peak at $H_{\pkb}$. Owing to the interfacial
origin of the voltage-FMR signal, the peak at $H_{\pkc}$ is more
distinguishable in the spectrum of the voltage-FMR (not shown), and
its linewidth, $\Delta H_{\pkc}=70$~Oe, can be fitted. It corresponds
to a damping $\alpha_{\pkc}=1.29 \times 10^{-2}$.

The linewidths of the modes at $H_{\pkA}$ and $H_{\pkC}$ can also be
fitted and give similar results for the damping associated to each
layer. In the case of the RF current excitation, a frequency-sweep
spectrum can be acquired at a fixed bias magnetic field $H_\text{fix}$
(see FIG.~\ref{fig:04}). In that case, the damping constant is simply
obtained by $\alpha=\Delta f/(2f)$, where $\Delta f$ is the width of
the line centered at $f$. At $H_\text{fix}=10$~kOe,
$f_{\pkA}=7.37$~GHz and $\Delta f_{\pkA}=0.12$~GHz, which yield
$\alpha_{\pkA}=0.81 \times 10^{-2}$, and $f_{\pkC}=10.92$~GHz and
$\Delta f_{\pkC}=0.33$~GHz, which yield $\alpha_{\pkC}=1.5 \times
10^{-2}$.

In summary, we retain the following values for the damping parameters
in respectively the thin and the thick layers: $\alpha_{a}=(1.4\pm0.2)
\times 10^{-2}$ and $\alpha_{b}=(0.85\pm0.1) \times 10^{-2}$.  We have
reported them, together with $\gamma$, in Table~\ref{tab:param}.

These two values are in line with the ones obtained on the reference
film, which have also been reported in Table~\ref{tab:param}. Still,
we observe that the linewidths in the nanostructure are systematically
lower than the ones measured on the reference film. This is a constant
characteristic that we associate to the confined geometry, which lifts
most of the degeneracy (well separated SW modes) and thus strongly
reduces the inhomogeneous part of the linewidth observed in the
infinite layer \cite{loubens07,chen08}. Rather, the inhomogeneities
associated to the magnetic layers \cite{loubens07} or to the
confinement geometry will lead to some mode splitting in the
nanostructure (see section~\ref{sec:asym}). We have checked that the
inhomogeneous contribution to the linewidth in the nano-pillar is
weak, by following the dependence of the measured $\Delta H$ as a
function of frequency. In fact, the increase of $\Delta H_{\pkc}$ from
70~Oe at 8.1~GHz to 105~Oe at 11~GHz is purely homogeneous.

Finally, the finding that the damping is larger in the thin layer than
in the thick layer is ascribed to the adjacent metallic layers
\cite{mizukami01}. In fact, non-local effects such as the spin pumping
effect \cite{tserkovnyak05,beaujour06a} and the spin diffusion in the
adjacent normal layers by the conduction electrons yield an
interfacial increase of the magnetic damping \cite{hurdequint07},
stronger in the case of thin layers.

\begin{table}
  \caption{Magnetic parameters of the thin Py$_a$ and thick Py$_b$ layers
    measured by cavity-FMR on the reference film (top row) and by
    mechanical-FMR in the nano-pillar (bottom row).}
  \begin{ruledtabular}
    \begin{tabular}{c c c c c}
      $4 \pi M_a$ (G) &  $\alpha_a$  & $4 \pi M_b$ (G) &
      $\alpha_b$  & $\gamma$ (rad $\cdot$ s$^{-1}$ $\cdot$ G$^{-1}$) \\
      \hline \\
      $8.2 \times 10^{3}$ & $1.5 \times 10^{-2}$ & $9.6 \times 10^{3}$ & 
      $0.9 \times 10^{-2}$  & $1.87\times 10^{7}$  \\
      \hline\\
      $8.0 \times 10^{3}$ & $1.4 \times 10^{-2}$ & $9.6 \times 10^{3}$ & 
      $0.85 \times 10^{-2}$  & $1.87\times 10^{7}$  \\
    \end{tabular}
  \end{ruledtabular}\label{tab:param}
\end{table}

\section{Theoretical analysis  \label{sec:theory}}

In this section, we first review a general formalism allowing the
calculation of the discrete spectrum associated with SW propagation
inside a confined body of arbitrary magnetic configuration. It is
shown that in the two-dimensional (2D) axially symmetric case,
different $\ell$-index modes can be excited separately, as found
experimentally in section~\ref{sec:hrfvsirf}. The classification of
the SW modes in this case is also used to extract the parameters of
each magnetic layer from the experimental FMR spectra. In a second
part, we discuss the influence of the dynamic coupling between the
magnetic disks, where the collective dynamics splits into binding and
anti-binding modes. It is shown that in our experimental case, the
dynamic dipolar coupling introduces a weak spectral shift, although
its influence on the character of the SW modes is real. In the last
part, a comparison to full three-dimensional (3D) micromagnetic
simulations is performed in order to study in details the collective
dynamics in the nano-pillar.

\subsection{Analytical model \label{sec:analytical}}

\subsubsection{General theory \label{sec:general}}

\begin{figure}
  \includegraphics[width=8.5cm]{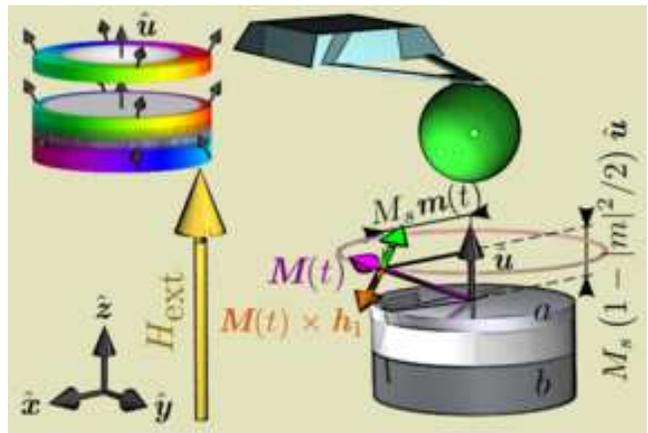}
  \caption{(Color online) Schematic representation of the
    magnetization dynamics under continuous RF excitation. In the
    steady state, the torque exerted by the RF perturbation field
    $\mathbi{h}_1$ (orange arrow) compensates the torque exerted by
    the damping (green), and the local magnetization vector
    $\mathbi{M}(t)$ (purple) precesses at the Larmor frequency on a
    circular orbit around the local equilibrium direction (unit vector
    $\unit u$). $\mathbi{M}(t)$ is the vector sum of a small
    oscillating component $M_s \bm m$ and a large static component
    $M_s \left( 1 - |\bm m|^2/2 \right)$, respectively transverse and
    parallel to $\unit u$. The inset shows the simulated spatial
    distribution of $\unit u$ inside the nano-pillar at
    $\mathbi{H}_{\rm ext}=10$~kOe (see section~\ref{sec:simu}). In the
    white regions, the magnetization is aligned along the normal
    $\unit z$ within 0.05$^\circ$. In the colored regions, $\unit u$
    is flaring ($<5^\circ$) in the radial direction (the hue indicates
    the direction of $\unit u - \unit z$ according to the color code
    defined in FIG.~\ref{fig:06}).}
  \label{fig:05}
\end{figure}

Below, we briefly review the general theory of linear SW excitations
(see appendix~\ref{app:motion} for more details). We consider an
arbitrary equilibrium magnetic configuration, where the local
magnetization writes $M_s\unit u$, with $M_s$ the saturation
magnetization and $\unit u$ the unit vector along the local
equilibrium direction (implicitly dependent on the spatial
coordinates). The linearization of the \emph{local} equation of motion
is obtained by decomposing the instantaneous magnetization vector $\bm
M(t)$ into a static and dynamic component \cite{gurevich96} (see
FIG.~\ref{fig:05}). We shall use the following ansatz:
\begin{equation}\label{eq:m-ansatz}
  \frac{\bm M(t)}{M_s} = \unit u + \bm m(t) + {\cal O}(\bm m^2)\,,
\end{equation}
where the transverse component $\bm m(t)$ is the small dimensionless
deviation ($ |\bm m| \ll 1$) of the magnetization from the equilibrium
direction. In ferromagnets, $|\bm M|=M_s$ is a constant of the motion,
so that the local orthogonality condition $\unit u \cdot \bm m = 0$ is
required.

Substituting Eq.~(\ref{eq:m-ansatz}) in the lossless Landau-Lifshitz
equation Eq.~(\ref{LL-pert}) (see appendix~\ref{app:motion}) and
keeping only the terms linear in $\bm m$, one obtains the following
dynamical equation for $\bm m$:
\begin{equation}\label{eq:m}
  \frac{\partial \bm m}{\partial t}= \unit u \times \op\Omega \ast \bm m \,,
\end{equation}
where here and henceforth, tensor operators are indicated by wide hat,
the cross product is denoted by $\times$ and the convolution product
is denoted by $\ast$. The self-adjoint tensor operator $\op\Omega$
represents the Larmor frequency:
\begin{equation}\label{eq:op-L}
  \op\Omega = \gamma H \op I + 4 \pi \gamma M_s \op G\, ,
\end{equation}
where $\gamma$ is the modulus of the gyromagnetic ratio, $H$ is the
scalar effective magnetic field, $\op I$ is the identity matrix, and
$\op G$ is the linear tensor operator describing the magnetic
self-interactions. The later is the addition of several contributions
$\op G^{(d)} + \op G^{(e)} + ...$, respectively the magneto-dipolar
interactions, the inhomogeneous exchange, etc... (see
appendix~\ref{app:na}). The effective magnetic field $\bm H$ is a
vector aligned along $\unit u$, whose norm is
\begin{equation}\label{eq:heff}
  H  = \unit u \cdot \bm H_0 - 4 \pi M_s \unit u \cdot \op G \ast
  \unit u\, , 
\end{equation}
the sum of the $\unit u-$component of $\bm H_0$, the total applied
magnetic field including the stray field of any nearby magnetic object
(in our case, the adjacent magnetic layer in the nano-pillar and the
spherical probe), reduced by the static self-interactions, which
include the depolarization magnetic field along $\unit u$ created by
the static component of the magnetization.

SW modes $\bm m_\nu$ are by definition eigen-solutions of
Eq.~(\ref{eq:m}):
\begin{equation}\label{eq:m-eigen}
  -i\omega_\nu\bm m_\nu= \unit u \times \op\Omega \ast \bm m_\nu \,.
\end{equation}
Here $\omega_\nu$ is the SW eigen-frequency and $\nu$ is a set of
indices to enumerate the different modes.

The main properties of SW excitations follow from the eigen problem
Eq.~(\ref{eq:m-eigen}) and the fact that the operator $\op\Omega$ is
self-adjoint and real. One can show that the eigen solutions obey the
closure relation
\begin{equation}\label{ortho}
  i \<\cc m_\nu \cdot (\unit u \times \bm m_{\nu'}) \> =  \mathcal{N}_\nu \delta_{\nu,\nu'}\, ,
\end{equation}
where $\delta$ is the Kronecker delta function and $\cc m$ stands for
the complex conjugate of $\bm m$. Here we have used the chevron
bracket notation introduced in Eq.~(\ref{eq:average}) to denote the
spatial average. The quantities $\mathcal{N}_\nu$ are real
normalization constants, which depend on the choice of eigen-functions
$\bm m_\nu$. If the equilibrium magnetization $\unit u$ corresponds to
a (local) minimum of the energy, then the operator $\op\Omega$ is
positive-definite. It follows that the ``physical'' modes with
$\omega_\nu >0$ have positive norm $\mathcal{N}_\nu > 0$. In this
formalism, the eigen-frequencies $\omega_\nu$ can be calculated as
\begin{equation}\label{omega-integral}
  \omega_\nu   = \frac{\<\cc m_\nu \cdot \op\Omega \ast \bm
    m_\nu\>}{\mathcal{N}_\nu}.
\end{equation}
The importance of this relation is that the frequencies $\omega_\nu$
calculated using Eq.~(\ref{omega-integral}) are variationally stable
with respect to perturbations of the mode profile $\bm m_\nu$. Thus,
injecting some trial vectors inside Eq.~(\ref{omega-integral}) allows
one to get approximate values of $\omega_\nu$ with high accuracy
\cite{bailleul06}. The trial vectors should obey some simple
properties: i) they should form a complete basis in the space of
vector functions $\bm m$, ii) be locally orthogonal to $\unit u$ and
iii) satisfy appropriate boundary conditions at the edges of the
magnetic body \cite{guslienko02}.

\subsubsection{Normally magnetized disks \label{sec:normal}}

In this part, we shall establish a SW modes basis $\bm m_\nu$ for a
normally magnetized disk. A specific feature of the considered
geometry is its azimuthal symmetry. Mathematically, this means that
the operator $\unit u \times \op\Omega$ commutes with the operator
$\op R_z$ that describes an infinitesimal rotation about the $\unit z$
axis, assuming that the boundary conditions are invariant under such a
rotation.

This particular configuration allows us to classify the SW modes
according to their behavior under the rotations in the $(x,y)$
plane. Namely, SW eigen-modes are also eigen-functions of the operator
$\op R_z$ corresponding to a certain integer azimuthal number $\ell$:
\begin{equation}\label{ell}
  \frac{\partial\bm m}{\partial\phi} - \unit z \times \bm m = -i(\ell-1)\bm m.
\end{equation}
Here, $\phi$ is the azimuthal angle of the polar coordinate system.

As one can see, Eq.~(\ref{ell}) determines the vector structure of
SW modes and their dependence on the angle $\phi$. Namely,
Eq.~(\ref{ell}) for a fixed $\ell$ has two classes of solutions:
\begin{subequations}\label{ell-sols}
\begin{equation}\label{ell-sols-a}
	\bm m_\ell^{(1)} = \frac 12 (\unit x + i \unit y)e^{-i\ell\phi}\psi_\ell^{(1)}(\rho)
\,,\end{equation}
and
\begin{equation}\label{ell-sols-b}
	\bm m_\ell^{(2)} = \frac 12 (\unit x - i \unit y)e^{-i(\ell-2)\phi}\psi_\ell^{(2)}(\rho)
\,,\end{equation}
\end{subequations}
where the functions $\psi_\ell^{(1,2)}(\rho)$ describe the dependence
of the SW mode on the radial coordinate $\rho$ and have to be
determined from the dynamical equations of motion. So, the azimuthal
symmetry allows one to reduce the 2D ($\rho$ and $\phi$) vector
equations to a one-dimensional ($\rho$) scalar problem.

Generally speaking, SW eigen-modes are certain linear combinations of
both possible $\ell$-forms Eqs.~(\ref{ell-sols}). The coupling of
these two forms is due solely to the inhomogeneous dipolar
interaction. In our experimental case (lowest energy modes of a
relatively thin disk) one can completely neglect this coupling
\cite{footnoteVasyl} and consider only the right-polarized form
Eq.~(\ref{ell-sols-a}). In the following we will drop the superscript
$(1)$ in $\bm m_\ell^{(1)}$ and $\psi_\ell^{(1)}$.

We shall now find an appropriate set of radial functions
$\psi_\ell(\rho)$ to calculate the SW spectrum using
Eq.~(\ref{omega-integral}). Here, we can take advantage of the
variational stability of Eq.~(\ref{omega-integral}) and, instead of
the exact radial profiles $\psi_\ell(\rho)$ (to find them one has to
solve integro-differential equations), use some reasonable set of
functions. Namely, it is known that the dipolar interaction in thin
disks or prisms does not change qualitatively the profile of SW modes,
but introduces effective pinning at the lateral boundaries
\cite{guslienko02}. Therefore, we will use radial profiles of the form
$\psi_\ell(\rho) = J_\ell(k_{\ell,n}\rho)$, where $J_\ell(x)$ is the
Bessel function and $k_{\ell,n}$ are SW wave-numbers determined from
the pinning conditions at the disk boundary $\rho = R$. For our
experimental conditions ($t_a,t_b\ll R$), the pinning is almost
complete, and we shall use $k_{\ell,n} = \kappa_{\ell,n}/R$, where
$\kappa_{\ell,n}$ is the $n$-th root of the Bessel function of the
$\ell$-th order.

\begin{figure}
  \includegraphics[width=8.5cm]{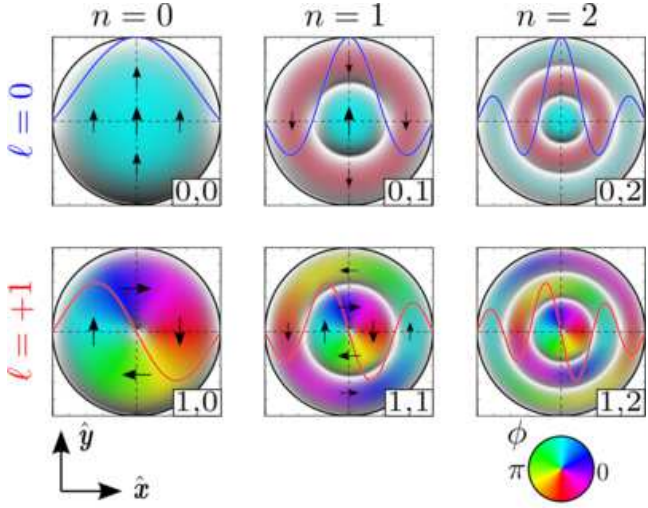}
  \caption{(Color online) Color representation of the Bessel spatial
    patterns for different values of the azimuthal mode index $\ell$
    (by row) and radial mode index $n$ (by column). The arrows are a
    snapshot of the transverse magnetization $\bm m_\nu$, labeled by
    the index $\nu = \ell,n$. All arrows are rotating synchronously
    in-plane at the SW eigen-frequency. In our coding scheme, the hue
    indicates the phase $\phi=\arg (\bm m_\nu) $ (or direction) of
    $\bm m_\nu$, and the brightness the amplitude of $|\bm
    m_\nu|^2$. The nodal positions ($|\bm m_\nu| =0$) are marked in
    white.}
  \label{fig:06}
\end{figure}

FIG.~\ref{fig:06} shows a color representation of the Bessel spatial
patterns for different values of the index $\nu=\ell,n$. We restrict
the number of panels to two values of the azimuthal mode index,
$\ell=0,+1$, with the radial index varying between $n=0,1,2$. In our
color code, the hue indicates the phase (or direction) of the
transverse component $\bm m_\nu$, while the brightness indicates the
amplitude of $|\bm m_\nu|^2$. The nodal positions are marked in
white. A node is a location where the transverse component vanishes,
\textit{i.e.}, the magnetization vector is aligned along the
equilibrium axis. This coding scheme provides a distinct visualization
of the phase and amplitude of the precession profiles. The black
arrows are a snapshot of the $\bm m_\nu$ vectors in the disk and are
all rotating synchronously in-plane at the SW eigen-frequency.

The top left panel shows the $\nu=0,0$ ($\ell=0$, $n=0$) mode, also
called the uniform mode. It usually corresponds to the lowest energy
mode since all the vectors are pointing in the same direction at all
time. Below is the $\ell=+1$, $n=0$ mode. It corresponds to SWs that
are rotating around the disk in the same direction as the Larmor
precession. The corresponding phase is in quadrature between two
orthogonal positions and this mode has a node at the center of the
disk. The variation upon the $n=0,1,2$ index ($\ell$ being fixed)
shows higher order modes with an increasing number of nodal
rings. Each ring separates regions of opposite phase along the radial
direction. All these spatial patterns preserve the rotation invariance
symmetry.

\subsubsection{Selection rules \label{sec:rules}}

\begin{figure}
  \includegraphics[width=8.5cm]{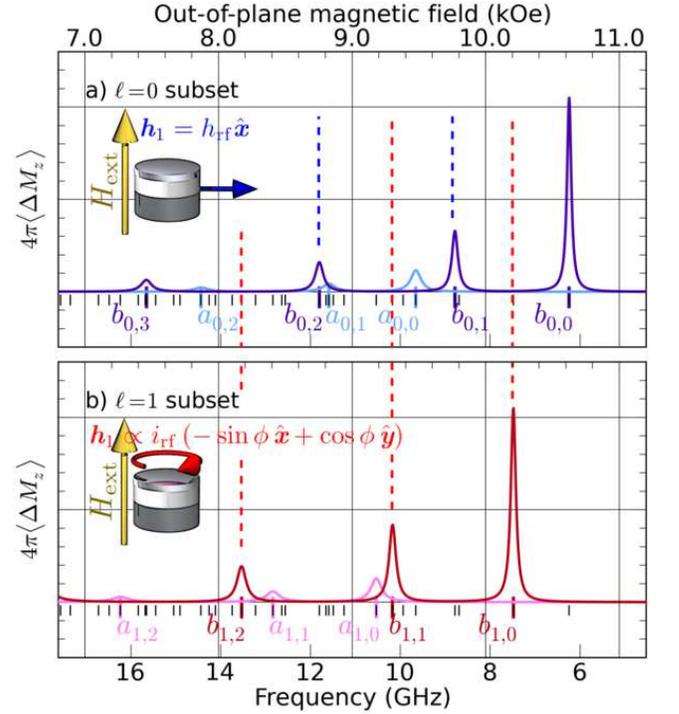}
  \caption{(Color online) Analytically calculated spectra at
    $H_\text{fix} = 10$~kOe using the set of Bessel functions (see
    FIG.~\ref{fig:06}) as the trial eigen-vectors. The panel (a) shows
    the linear response to a uniform excitation field $\unit h_1 =
    \unit x$ and the panel (b) to an orthoradial excitation field $\unit
    h_1 = -\sin\phi\,\unit x + \cos\phi\,\unit y$. A light
    (dark) color is used to indicate the energy stored
    Eq.~(\ref{eq:dmz}) in the thin Py$_a$ and thick Py$_b$ layers.}
  \label{fig:07}
\end{figure}

Using the complete set of Bessel functions in
Eq.~(\ref{omega-integral}), one can obtain analytically the discrete
spectrum of eigen-values for both the thin and thick layers. The
details of the numerical application can be found in
appendix~\ref{app:na}. The spectral values are displayed in
FIG.~\ref{fig:07} using vertical ticks labeled $\nu = j_{\ell n}$, where
$j=a,b$ indicates the precessing layer, and $\ell$, $n$ the azimuthal
and radial mode indices. They are calculated at fixed applied field
$H_\text{fix}=10$~kOe and placed on the graphs according to the
frequency scale below FIG.~\ref{fig:07}b, which is in correspondence
with the field scale above FIG.~\ref{fig:07}a (see \ref{sec:gyro} for
the equivalence between field- and frequency-sweep experiments).

The comparison with the experimental data in FIGS.~\ref{fig:02}a and
\ref{fig:02}b shows that the coupling to an external coherent source
depends primarily on the $\ell$-index. Indeed, this index carries the
discriminating symmetry in SW spectroscopy \cite{dillon60}. This is
because the excitation efficiency is proportional to the overlap
integral
\begin{equation}\label{eq:hnu}
  h_{\nu} =\frac{\< \cc m_\nu \cdot \bm h_1\>}{\mathcal{N}_\nu} \,,
\end{equation}
where $\bm h_1(\bm r)$ is the spatial profile of the external
excitation field. It can be easily shown that a uniform RF magnetic
field, $\bm h_1 = h_\text{rf} \mathbi{x}$, can only excite $\ell=0$ SW
modes. We have shown in FIG.~\ref{fig:07}a the predicted position of
these modes with blue tone ticks. Obviously the largest overlap is
obtained with the so-called uniform mode ($n=0$). Higher radial index
modes ($n\neq0$) still couple to the uniform excitation but with a
strength that decreases as $n$ increases
\cite{damon61,charbois02a}. The $\ell\neq 0$ normal modes, however,
are hidden because they have strictly no overlap with the
excitation. The comparison with the experimental spectrum in
FIG.~\ref{fig:02}a confirms that conventional FMR \cite{kittel58}
probes only partially the possible SW eigen-modes, along the
$\ell=0$-index value. In contrast, the RF current-created Oersted
field, $\bm h_1 = h_\text{Oe}(\rho) (-\sin\phi\,\unit x +
\cos\phi\,\unit y)$ has an orthoradial symmetry and can only excite
$\ell = +1$ SW modes. We have shown in FIG.~\ref{fig:07}b the
predicted position of these modes with red tone ticks. They are in
good agreement with the resonance positions observed in
FIG.~\ref{fig:02}b. We also note that the $\ell=0$ and $\ell=+1$
spectra calculated analytically bear similar $a/b$ and $n$ index
series as a function of energy. This explains why the two spectra in
FIGS.~\ref{fig:02}a and \ref{fig:02}b look in translational
correspondence with each other. We emphasize that the same
translational correspondence would have been observed for any higher
azimuthal order $\ell > 1$ index spectra.

From the coupling to the excitation field expressed by
Eq.~(\ref{eq:hnu}), one can also calculate the mechanical-FMR signal
$\propto \langle \Delta M_z \rangle$, proportional to the energy
stored in the magnetic system \cite{klein03,loubens05}. For an
arbitrary pulsation frequency $\omega$,
\begin{equation} \label{eq:dmz} 4 \pi \< \Delta \bm M \cdot \unit u\>
  \simeq 4 \pi M_s \sum_\nu \frac{\gamma^2 | h_\nu |^2
  }{(\omega-\omega_\nu)^2 + \Gamma_\nu^2} \mathcal{N}_\nu\,,
\end{equation}
where the SW damping rate $\Gamma_\nu$ is given by
Eq.~(\ref{eq:gamma}) in appendix~\ref{app:motion}. Eq.~(\ref{eq:dmz})
is derived under the approximation that the only relevant coefficients
in the damping matrix are the diagonal terms. It has been used to
compute the relative peak amplitudes in the analytically calculated
spectra of FIG.~\ref{fig:07}.

\subsubsection{Comparison with experiments \label{sec:compar}}

The analytical model outlined in sections~\ref{sec:general} and
\ref{sec:normal} can be used to analyze the experimental spectra of
FIG.~\ref{fig:02}, and to extract some useful parameters of the
nano-pillar. More details can be found in the appendix~\ref{app:na}
along with an approximate expression for the SW frequencies in the
form of Kittel’s traditional formula (with renormalized values of the
effective self-demagnetization fields). This Kittel’s formula, derived
for the $\ell=0$ spectrum, should be used to analyze the SW spectrum
excited by a uniform RF field to yield the correct values of the
magnetization in our nano-pillar. Identifying the experimental peaks
at $H_{\pkc}$ and $H_{\pka}$ as the lowest energy modes of the thin
Py$_a$ and thick Py$_b$ layers yields their respective magnetizations
$4\pi M_a=8.0 \times 10^3$~G and $4 \pi M_b=9.6 \times 10^3$~G, see
Eq.~(\ref{eq:omega}). These values have been reported in
Table~\ref{tab:param}, together with those measured in the reference
film (see appendix~\ref{app:cavity}). The magnetizations extracted in
the nano-pillar are the same as in the extended film. The only small
difference concerns the magnetization of the thin layer, which is
200~G lower in the nanostructure than in the reference film (where
$4\pi M_a=8.2 \times 10^3$~G). We attribute this to some
interdiffusion between Py and Cu or Au at the interfaces of the thin
layer, which can happen during the etching process of the nano-pillar.

Second, the separation between SW modes crucially depends on the
lateral confinement in the nano-pillar and thus on the precise value
of its radius. Experimentally, the measured field separation between
the two first peaks in FIG.~\ref{fig:02}a (FIG.~\ref{fig:02}b), which
differ by an additional node in the radial direction, is
$H_{\pka}-H_{\pkb}=1.04$~kOe ($H_{\pkA}-H_{\pkB}=1.05$~kOe). Using the
nominal radius $100$~nm in the analytical model predicts that
consecutive $n$-index mode ($n=0$ and $n=1$ modes) should be separated
by 1.33~kOe, which is larger than the observed value. This separation
drops to 1.05~kOe for a larger disk radius $R=125$~nm, which we thus
refer to as the radius of our nano-pillar. This value of $R$ also
allows to estimate the shift between the $\ell=0$ and $\ell=+1$
spectra, found to be 530~Oe, in good agreement with the experimental
value $H_{\pka}-H_{\pkA}=470$~Oe observed in FIG.~\ref{fig:02}.

\subsection{Influence of dipolar coupling between different
  layers \label{sec:dipolar}}

\begin{figure}
  \includegraphics[width=8.5cm]{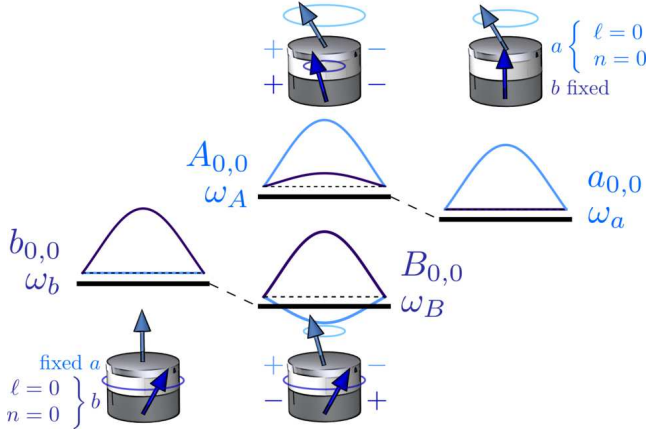}
  \caption{(Color online) Schematic representation of the coupled
    dynamics between two different magnetic disks. Here, $\omega_b$,
    the eigen-frequency of the lowest energy precession mode in the
    thick layer (the thin layer being fixed at equilibrium) is smaller
    than $\omega_a$, the one in the thin layer (the thick layer being
    fixed at equilibrium). When the two disks are dynamically coupled
    through the dipolar interaction, the binding state $B$ corresponds
    to the two layers oscillating in anti-phase at $\omega_B$, with
    the precession occurring mostly in the thick layer, whereas the
    anti-binding state $A$ corresponds to the layers oscillating in
    phase at $\omega_A$, with the precession mostly in the thin
    layer. This is shown by displaying the dipolar charges and the
    precession profile $\bm m(\rho)$ in each layer using a light
    (dark) color to represent the contribution of the thin (thick)
    layer.}
  \label{fig:08}
\end{figure}

In the treatment above we have neglected the dynamic coupling between
the two magnetic disks in dipolar interaction. In general, the
interaction between two identical magnetic layers will lead to the
hybridization of the same $\nu$-index mode of each layer into two
collective modes: the acoustic mode, where the layers are precessing
in phase, and the optical mode, where they are precessing in
anti-phase. This has been observed in interlayer-exchange-coupled thin
films \cite{belmeguenai07} and in trilayered wires where the two
magnetic stripes are dipolarly coupled \cite{gubbiotti04}. In the case
where the two magnetic layers are not identical (different geometry or
magnetic parameters), this general picture continues to
subsist. Although both isolated layers have eigen-modes with different
eigen-frequencies, the collective magnetization dynamics still splits
in a binding and anti-binding state. But here, the precession of
magnetization can be more intense in one of the two layers and the
spectral shift of the coupled SW modes with respect to the isolated SW
modes is reduced, as it was observed in both the
dipolarly-\cite{gubbiotti04} and exchange-coupled cases
\cite{benyoussef10}.

Here, we assume that the dominant coupling mechanism between the Py
layers is the magnetic dipolar interaction. We neglect any exchange
coupling between the magnetic layers mediated through the normal
spacer or any coupling associated to pure spin currents
\cite{woltersdorf07} in our all-metallic spin-valve structure. To
analyze the influence of the dipolar coupling between the two magnetic
layers, one can complement the perturbation theory derived in the
previous section~\ref{sec:analytical} and in the
appendix~\ref{app:motion}. Denoting $c_j$, the SW amplitudes in $j$-th
disk , one can get from Eq.~(\ref{eq:dcdt-1}):
\begin{subequations}\label{tmp}
\begin{eqnarray}
  \frac{dc_a}{dt} &=& -i\omega_ac_a + i\gamma h_{a,b}c_b
  \,,\\
  \frac{dc_b}{dt} &=& -i\omega_bc_b + i\gamma h_{b,a}c_a
  \,,
\end{eqnarray}
\end{subequations}
where $\omega_j$ is the frequency of the $j$-th disk ($j = a, b$) with
account of only the static field of the ${j^\prime}$-th disk
($j^\prime = b, a$) (\textit{i.e.}, with $\bm M_j^\prime$ fixed at
equilibrium, see FIG.~\ref{fig:08}). The cross term $h_{j,j^\prime}$ is
given by
\begin{equation} \label{eq:hjk} h_{j,j^\prime} = -\frac{4 \pi
    M_{j^\prime}}{\mathcal{N}_j}\< \cc m_j \cdot \op G^{(d)} \ast \bm
  m_{j^\prime}\>_j \ .
\end{equation} 
Here, $\op G^{(d)}$ represents the magneto-dipolar interaction,
$M_{j^\prime}$ is the saturation magnetization of the $j^\prime$-th
disk and the averaging goes over the volume of $j$-th disk. Thus,
$h_{j,j^\prime}$ is the average over the $j$-th mode of the magnetic
field created by the magnetization of the $j^\prime$-th disk. It can
be shown that the overlap defined in Eq.~(\ref{eq:hjk}) is maximum
between mode pairs bearing similar wave-numbers in each layer
(\textit{i.e.}, the same set of indices $\nu$) \cite{gubbiotti04}.
This is the reason why dropping the index $\nu$ in Eqs.~(\ref{tmp})
and (\ref{eq:hjk}) is a reasonable approximation.

The anti-binding ($A$) and binding ($B$) eigen-frequencies of
Eqs.~(\ref{tmp}) have the form
\begin{equation} \label{eq:AB} \omega_{A,B} =
  \frac{\omega_a+\omega_b}{2} \pm
  \sqrt{\left(\frac{\omega_a-\omega_b}{2}\right)^2+\Omega^2}
  \,,\end{equation} where
\begin{equation}
	\Omega^2 = \gamma^2 h_{a,b}h_{b,a}
\ .\end{equation}

In the case when the dipolar coupling is small ($\Omega \ll |\omega_a
- \omega_b|$), the eigen-frequencies can be written as (we assume
$\omega_a > \omega_b$)
\begin{eqnarray}
  \omega_A &=& \omega_a + \frac{\Omega^2}{\omega_a - \omega_b}
  \,,\\
  \omega_B &=& \omega_b - \frac{\Omega^2}{\omega_a - \omega_b}
  \ .
\end{eqnarray}
These equations can be used for quantitative purposes when
$\Omega/|\omega_a - \omega_b| < 0.3$ in which case they describe
frequency shift with accuracy better than 10\%. Thus, the larger of the
frequencies ($\omega_a$) shifts up by
\begin{equation}\label{eq:shift}
	\Delta\omega = \frac{\Omega^2}{\omega_a - \omega_b}
\,,\end{equation}
while the smaller one ($\omega_b$) shifts down by the same amount.
This effect is summarized in FIG.~\ref{fig:08}.

A numerical estimate of the coupling strengths $h_{a,b}$ and $h_{b,a}$
between the lowest energy SW modes in each disk can be found in
appendix~\ref{app:na}. The obtained result is very close to the
approximate estimation used in Ref.\cite{dmytriiev10}, where the
spatial structure of the interacting SW modes is ignored to calculate
the dipolar coupling between uniformly precessing disks. For the
experimental parameters, $\Omega/2\pi \simeq 0.5$~GHz. This coupling
is almost an order of magnitude smaller than the frequency splitting
$\omega_a-\omega_b$, caused, mainly, by the difference of effective
magnetizations of two disks: $\gamma 4 \pi (M_b - M_a) \simeq 2\pi
\cdot 4.5$~GHz. As a result, the shift of the resonance frequencies
due to the dipolar coupling is negligible, $\Delta\omega/2\pi \simeq
0.06$~GHz.

Using Eqs.~(\ref{tmp}), one can also estimate the level of mode
hybridization due to the dipolar coupling. For instance, at the
frequency $\omega_A \approx \omega_a$, the ratio between the
precession amplitudes in the two layers is given by
\begin{equation}\label{eq:ratio}
  \left|c_b/c_a\right|_{\omega_A} = \Delta\omega/(\gamma h_{a,b}) 
  \simeq \frac{\Omega}{\omega_a-\omega_b}\, .
\end{equation}
For the experimental parameters, $\Omega/(\omega_a-\omega_b) \approx
0.1$, \textit{i.e.}, the precession amplitude in the disk $b$ is about
$10~\%$ of that in the disk $a$. Thus, although the dipolar coupling
induces a small spectral shift (second order in the coupling
parameter, Eq.~(\ref{eq:shift})), its influence in the relative
precession amplitude is significant (first order in the coupling
parameter, Eq.~(\ref{eq:ratio})). Finally, we point that here the
dipolar coupling is anti-ferromagnetic, and that the binding (lower
energy) mode $B$ always corresponds to the thick layer mainly precessing,
with the thin layer vibrating in anti-phase, and vice-versa for the
anti-binding (in-phase) mode $A$ (see dipolar charges in
FIG.~\ref{fig:08}).

\subsection{Micromagnetic simulations \label{sec:simu}}

\begin{figure}
  \includegraphics[width=8.5cm]{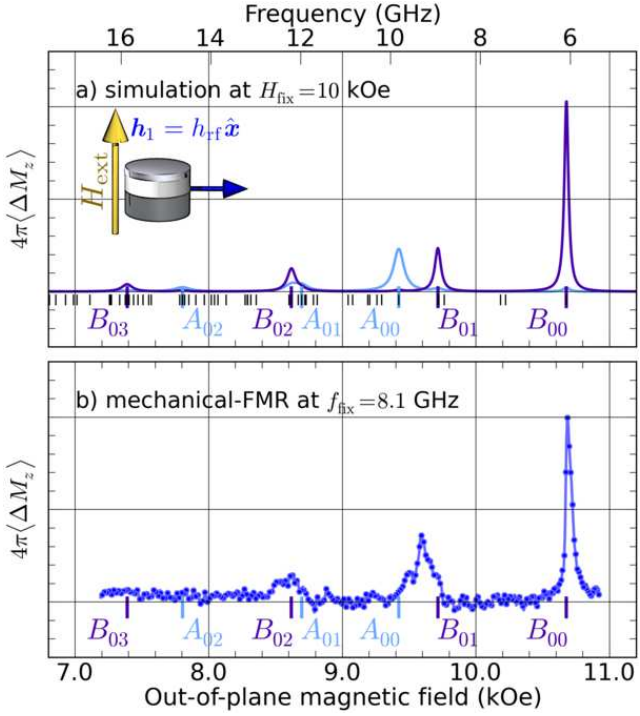}
  \caption{(Color online) Panel (a) is the numerically calculated
    spectral response to a uniform excitation field $\bm h_1 \propto
    \unit x$, from a 3D micromagnetic simulation performed at
    $H_\text{fix}=10$~kOe. The peaks are labeled according to their
    precession profiles shown in FIG.~\ref{fig:11}. A light (dark)
    color is used to indicate the energy stored in the thin (thick)
    layer. Panel (b) recalls the experimental spectrum measured by
    mechanical-FMR when exciting the nano-pillar by a homogeneous RF
    magnetic field at $f_\text{fix}=8.1$~GHz.}
  \label{fig:09}
\end{figure}

\begin{figure}
  \includegraphics[width=8.5cm]{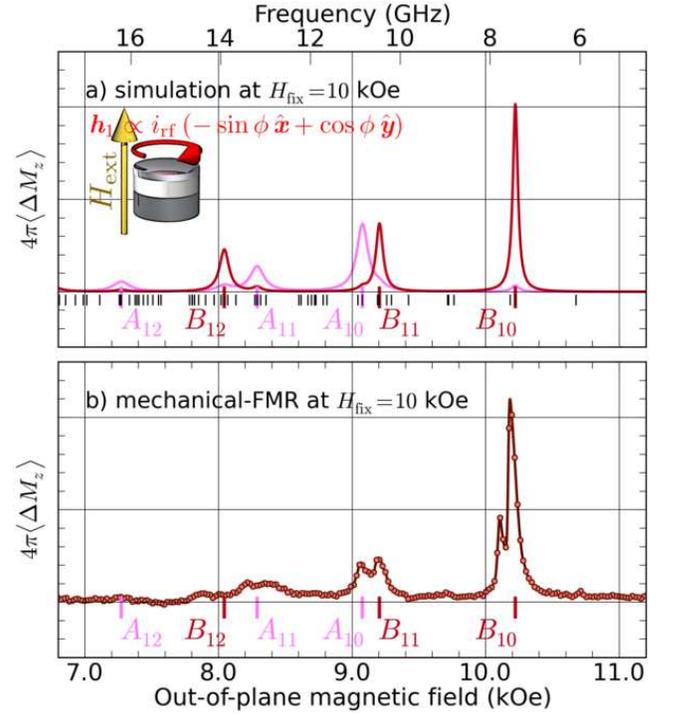}
  \caption{(Color online) Panel (a) is the simulated spectral response
    to an orthoradial excitation field $\bm h_1 \propto
    -\sin\phi\,\unit x + \cos\phi\,\unit y$. Panel (b) recalls the
    experimental spectrum measured by mechanical-FMR for an RF current
    excitation.}
  \label{fig:10}
\end{figure}

In the analytical formalism presented above, several approximations
have been made. For instance, we have assumed total pinning at the
disks boundary for the SW modes and no variation of the precession
profile along the disks thicknesses (2D model), and we have neglected
the dependence on $\nu$ of the dynamic dipolar coupling. Still, it
allows to extract important parameters in our nano-pillar, such as its
radius and the magnetization in both layers. It also describes the
influence of the dynamic dipolar coupling on the position and
collective character of the SW modes.

Instead of developing a more complex analytical formalism, we have
performed innovative 3D micromagnetic simulations in order to go
beyond the approximations mentioned above, and to unambiguously
identify the SW modes observed in our nano-pillar sample. For that
purpose, we have used a combination of micromagnetic simulation
solvers available as part of SpinFlow 3D, a finite element based
simulation platform for spintronics developed by In Silicio
\cite{SpinFlow3D}. The steady state micromagnetic solver used to
obtain numerical approximations of micromagnetic equilibrium states is
based on a weak formulation and Galerkin type finite element
implementation of the very efficient projection scheme introduced in
Ref.\cite{e01}. A second numerical solver, a micromagnetic Eigen
solver, has been used for fast calculations of lossless 3D SW
eigen-modes. It is based on a finite element discretization of the
generalized eigen-value problem defined by the linearized lossless
magnetization dynamics in the vicinity of an arbitrary pre-computed
equilibrium state, following an approach very similar to the one
introduced in Ref.\cite{aquino09}. The discrete generalized
eigen-value problem is solved with an iterative Arnoldi method using
the ARPACK library \cite{lehoucq98}. In this calculation the full
complexity of the 3D micromagnetic dynamics of the presently
considered bilayer system is preserved. The solver outputs both the
eigen-values by increasing energy order and the associated
eigen-vectors. Several tens of SW eigen-modes can be accurately
computed in a matter of few minutes of CPU time with a standard
desktop PC, for magnetic thin film nano-structures with typical
lateral sizes in the 100~nm range. This is two to three orders of
magnitude faster compared to the required computation time when using
more traditional approaches for micromagnetic computation of SW
eigen-modes, which are typically based on the Fourier component
analysis of time series generated by the solution of the full
non-linear Landau-Lifshitz-Gilbert equation
\cite{mcmichael05}. Finally, a quite generic linear response solver,
implementing among other things the spectral decomposition of the MRFM
signal as expressed in Eqs.~(\ref{eq:hnu}), (\ref{eq:dmz}) and
(\ref{eq:gamma}), has been used to compute the MRFM spectra shown
here.

To proceed, the nano-pillar is first discretized using unstructured
meshing algorithms resulting in an average mesh size of 3.5~nm. This
corresponds to a total number of vertices in the vicinity of $5 \times
10^{4}$. The magnetization vector is interpolated linearly inside each
cell (tetrahedra) -- a valid approximation taking into account that
the cell sizes are smaller than the exchange length $\Lambda_{\rm ex}
\simeq 5$~nm in Permalloy. The magnetic parameters introduced in the
code are the ones reported in Table~\ref{tab:param}, and the
simulation incorporates the perturbing presence of the magnetic sphere
attached on the cantilever. Moreover, the 10~nm thick Cu spacer is
replaced by vacuum, so that the layers are only coupled through the
dipolar interaction (spin diffusion effects are absent).

The next step is to calculate the equilibrium configuration in the
nano-pillar at $H_{\rm ext}=H_\text{fix}=10$~kOe. The external
magnetic field is applied exactly along $\unit z$ and the spherical
probe with a magnetic moment $m=2 \times 10^{-10}$~emu is placed on
the axial symmetry axis at a distance $s=1.3~\mu$m above the upper
surface of the nano-pillar. The convergence criterion introduced in
the code is $ | dM_z/M_j | < 2 \cdot 10^{-9}$ between iterations. The
result shown in the inset of FIG.~\ref{fig:05} reveals that the
equilibrium configuration is almost uniformly saturated along $\unit
z$. Still, a small tilt ($<5^\circ$) of the magnetization, away from
$\unit z$ and along the radial direction, is observed at the periphery
of the thick and thin layers.

\begin{figure}
  \includegraphics[width=8.5cm]{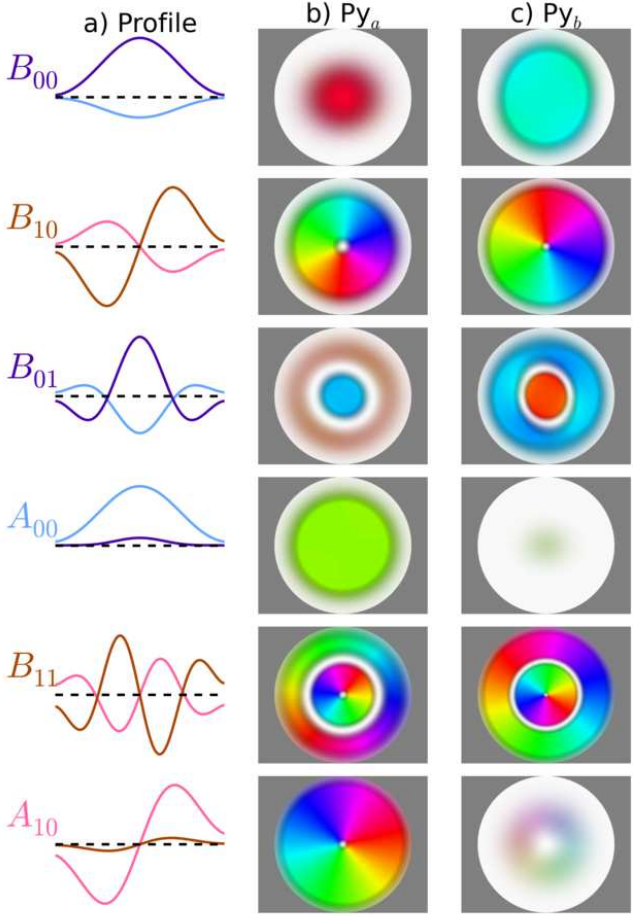}
  \caption{(Color online) Simulated precession patterns of the
    eigen-vectors. Column (a) shows the precession profiles across the
    thin (light color) and thick (dark color) layers. Columns (b) and
    (c) show the dynamics in the thin Py$_a$ and thick Py$_b$ layers,
    respectively, with the color code defined in FIG.~\ref{fig:06}.}
  \label{fig:11}
\end{figure}

The micromagnetic eigen solver is then used to compute the lowest
eigen-values of the problem as well as the associated
eigen-vectors. The discrete list of eigen-values under 18~GHz is shown
as black vertical ticks at the bottom of FIGS.~\ref{fig:09}a and
\ref{fig:10}a. The precession patterns of the six eigen-vectors
corresponding to the six lowest eigen-frequencies are shown in
FIG.~\ref{fig:11}. The middle and right columns show the dynamics $\bm
m$ in the thin Py$_a$ and thick Py$_b$ layers, while the precession
profiles along the median direction are shown on the left in light and
dark colors, respectively. The resonance peaks are labeled according
to the SW modes precession profiles and the eigen-values of the
simulated peaks are reported in Table~\ref{tab:eigen}.

From the eigen-vectors spatial patterns, one can compute their
coupling (Eq.~(\ref{eq:hnu})) to a uniform RF field $\bm h_1 =
h_\text{rf} \unit x$ and, with Eq.~(\ref{eq:dmz}), the mechanical-FMR
spectrum (FIG.~\ref{fig:09}a). The same procedure is repeated for the
RF current-induced Oersted field $\bm h_1 \propto i_\text{rf}
(-\sin\phi\,\unit x + \cos\phi\,\unit y)$ excitation
(FIG.~\ref{fig:10}a). Since the code gives access to the contribution
of each layer, a light (dark) tone is used to indicate the vibration
amplitude in the thin (thick) layer in the two figures. For
comparison, the mechanical-FMR spectra of FIGS.~\ref{fig:02}a and
\ref{fig:04}a have been reported in FIGS.~\ref{fig:09}b and
\ref{fig:10}b, respectively. We have applied the same conversion
between the frequency (top) and field (bottom) scales as discussed in
section~\ref{sec:gyro}.

\begin{table}
  \caption{Comparative table of the resonance values for the SW modes,
    arranged in order of increasing energy. On the left are the consecutive
    peak locations measured experimentally. Experiments are performed at
    $f_\text{fix}=8.1$~GHz (FIG.~\ref{fig:02}) or $H_\text{fix}=10$~kOe
    (FIG.~\ref{fig:04}a). On the right are the simulated eigen-frequencies $f$
    at $H_\text{fix}=10$~kOe. The conversion to field value $H_\text{ext}$ is obtained
    through $H_\text{ext} - H_\text{fix}=2\pi (f-f_\text{fix})/\gamma$.}
  \begin{ruledtabular}
    \begin{tabular}{c c c | c c c}
Exp. & $f$ (GHz) & $H_\text{ext}$ (kOe) & Simu. & $f$ (GHz) & $H_\text{ext}$ (kOe)\\
\hline
${\pka}$ &       & 10.69 & $B_{00}$ &  6.08 & 10.68 \\
${\pkA}$ &  7.37 & 10.22 & $B_{10}$ &  7.44 & 10.22 \\ 
${\pkb}$ &       & 9.65 & $B_{01}$ &  8.95 & 9.71 \\
${\pkc}$ &       & 9.51 & $A_{00}$ &  9.82 & 9.42 \\
${\pkB}$ & 10.48 & 9.17 & $B_{11}$ & 10.47 & 9.20 \\ 
${\pkC}$ & 10.92 & 9.07 & $A_{10}$ & 10.85 & 9.08 \\
${\pkd}$ &       & 8.64 & $A_{01}$ & 11.98 & 8.69 \\
${\pkD}$ & 13.41 & 8.22 & $A_{11}$ & 13.19 & 8.29 \\
    \end{tabular}
  \end{ruledtabular}\label{tab:eigen}
\end{table}

In FIG.~\ref{fig:09}a, the largest peak in the simulation occurs at
the same field as the experimental peak at $H_{\pka}$. This lowest
energy mode corresponds to the most uniform mode with the largest
wave-vector and no node along the radial direction, thus it has the
index $n=0$. It has uniform phase along the azimuthal direction, which
is the character of the $\ell=0$ index. For this mode, the thick layer
is mainly precessing, with the thin layer oscillating in anti-phase
(binding index $B$), as can be seen from its spatial profile in
FIG.~\ref{fig:11}. The same analysis can be made for the second peak,
labeled $B_{01}$, which occurs close to the peak at $H_{\pkb}$. It
also corresponds to a resonance mainly of the thick layer, and its
color representation shows that this is the first radial harmonic
($n=1$), with one line of nodes in the radial direction. Again, the
thin layer is oscillating in anti-phase, with the same radial index
$n=1$, as clearly shown by the mode profile along the median
direction. The third peak is labeled $A_{00}$ and is located close to
the experimental peak at $H_{\pkc}$. It corresponds this time to a
uniform ($n=0$) precession mainly located in the thin layer, in
agreement with the experimental analysis presented in
section~\ref{sec:idc}. In this mode, the thick layer is also vibrating
in phase with the thin layer (anti-binding index $A$).

We can also look at the relative amplitudes of precession in the two
disks to quantify the dynamic coupling between the disks. From the
profiles shown in FIG.~\ref{fig:11}, one can infer that for the
fundamental mode $B_{00}$, the amplitude of precession is distributed
with a ratio of about 3:1 between the thick (75\%) and the thin layer
(25\%). For the mode $A_{00}$, the ratio is 8:1 in favor of the thin
layer, which contributes to 89\% of the precession amplitude (11\% for
the thick layer). These relative precession amplitudes were expected
from the relative weight of the thick and thin layers and from the
approximate analytical model presented in
section~\ref{sec:dipolar}. The simulated field separation between the
two coupled uniform modes $(\omega_{B_{00}} - \omega_{A_{00}})/\gamma
= 1.28$~kOe compares also well with the 1.30~kOe estimate from the 2D
model, with the dynamic dipolar coupling taken into account. Finally,
one can check from the simulations the independence of the precession
profiles on the thickness. This confirms the validity of the 2D
approximation and explains the performances of the analytical model.

We now briefly comment on the simulated spectrum of FIG.~\ref{fig:10},
which enables to identify the SW modes excited by the orthoradial
Oersted field produced by the RF current flowing through the
nano-pillar. From FIGS.~\ref{fig:10} and \ref{fig:11}, it is clear
that the modes which couple to this excitation symmetry have a
rotating phase in the azimuthal direction, characteristic of the
$\ell=+1$ modes \cite{footnotemirror}. We find that the SW modes of
FIG.~\ref{fig:10} show the same series of $A/B$ and $n$ indices as
those in FIG.~\ref{fig:09} (but their $\ell$-index is different). This
sustains the translational correspondence between the SW spectra of
FIGS.~\ref{fig:02}a and \ref{fig:02}b. Finally, we point out that, for
all the modes displayed in FIG.~\ref{fig:11}, the pinning conditions
at the boundaries of each disk are not trivial, which we attribute to
the collective nature of the motion driven by the dipolar coupling
\cite{kostylev04}. The general trend observed here is that the thin
layer is less pinned than the thick layer for in-phase modes, and vice
versa.

To summarize, the 3D micromagnetic simulations enable the
identification (with three indices, $A/B$, $\ell$ and $n$) of the SW
modes probed experimentally by both a uniform RF magnetic field and an
RF current flowing through the nano-pillar, \textit{i.e.}, of their
respective selection rules. They confirm the experimental analysis
performed in section~\ref{sec:exp} and give a deeper insight on the
collective nature of the magnetization dynamics in the nano-pillar
discussed in section~\ref{sec:dipolar}.

\section{Symmetry breaking \label{sec:breaking}}

In the following, we review some characteristic spectral features
associated with the breaking of the axial symmetry in our
experiment. First, we experimentally report on the appearance of
$\ell=0$ modes in the SW spectrum excited by an RF current flowing
through the nano-pillar, when a small tilt angle is introduced between
the applied field and the normal of the layers. This bridges the gap
between our mechanical-FMR experiments and usual ST-FMR measurements
\cite{sankey06,chen08}. Second, we have simulated the spectral
distortions introduced by breaking the cylindrical symmetry of the SW
confinement potential. This enables to explain the lift of degeneracy
in the SW spectrum, which leads to the splitting of modes.

\subsection{Polar angle dependence \label{sec:polar}}

\begin{figure}
  \includegraphics[width=8.5cm]{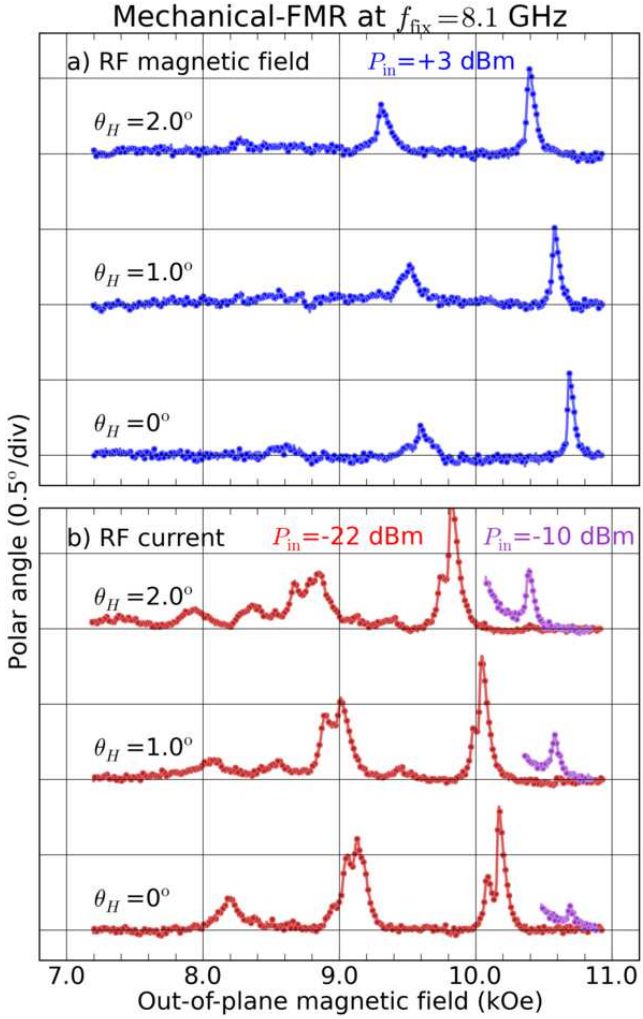}
  \caption{(Color online) Dependence of the mechanical-FMR spectra
    excited by a uniform RF magnetic field (a) and by an RF current
    flowing through the nano-pillar (b) on the polar angle $\theta_H$
    between the applied field and the normal to the layers. Superposed
    (in purple) is the behavior of the high field tail at larger
    power.}
  \label{fig:12}
\end{figure}

The dependence on the polar angle $\theta_H=(\unit z,\bm{H}_{\rm
  ext})$ of the mechanical-FMR spectra excited by a uniform RF
magnetic field and by an RF current flowing through the nano-pillar is
presented in FIG.~\ref{fig:12}. Let us first focus on the conventional
FMR spectra shown in FIG.~\ref{fig:12}a, acquired at three different
polar angles from the exact perpendicularity, increasing by steps of
$1^\circ$. The main effect here is the shift of the $\ell=0$ SW modes
spectrum towards lower field as $\theta_H$ increases, which has been
explained in details in Ref.\cite{klein08} for a single magnetic
disk. It is due to the decrease of the demagnetizing field produced by
the tilt of the equilibrium magnetization away from the normal. In
fact, in each magnetic layer $j=a,b$, the uniform magnetization
creates a non-uniform dipolar field $4 \pi M_j \op G^{(d_j)} \ast
\unit u_j$, which is maximum in the exact normal configuration. The
equilibrium direction $\unit u_j$ is in the plane $(\unit z,
\bm{H}_{\rm ext})$ and makes a polar angle $\theta_j>\theta_H$ with
the normal determined by Eq.~(\ref{eq:equil}). It can be estimated
that when $H_{\rm ext}\approx10$~kOe and $\theta_H$ increases from
$0^\circ$ to $2^\circ$, the equilibrium angles $\theta_a$ and
$\theta_b$ of the static magnetization in the thin and thick layers
linearly increases from $0^\circ$ to $\approx 9^\circ$ and from
$0^\circ$ to $\approx 13^\circ$, respectively. This leads to a shift
to lower field of the FMR spectrum by about $420$~Oe (see
appendix~\ref{app:na}), in agreement with the data. We also emphasize
that, in fact, the profiles of the SW eigen-modes are affected by the
breaking of axial symmetry, and that the pure $\ell=0$ eigen-modes
when $\theta_H = 0$ become mixed with $\ell\neq0$ modes \cite{klein08}
when $\theta_H \neq 0$.

We now turn to the influence of the polar angle $\theta_H$ on the FMR
spectra excited by an RF current ($i_\text{rf}=170~\mu$A). The same
global shift towards lower field as discussed above is observed in
FIG.~\ref{fig:12}b by looking at the red spectra acquired with an
increasing $\theta_H$. But there is an important additional effect
here. Whereas only $\ell=+1$ SW modes are excited by the RF current
flowing through the nano-pillar in the exact perpendicular geometry,
resonance peaks can also be detected at the positions of $\ell=0$ SW
modes when $\theta_H\neq 0$. Although the amplitudes of the $\ell=0$
modes are not large in FIG.~\ref{fig:12}b, it is quite clear that they
all grow as $\theta_H$ increases. In order to reveal this effect
better, we have reported in purple on the same figure the resonance
peak of the mode $B_{00}$ excited with a +12~dB larger power
($i_\text{rf} \simeq 680$~$\mu$A), as a function of
$\theta_H$. Despite the large RF current excitation, its amplitude
almost vanishes at $\theta_H=0$. Then, it increases linearly with
$\theta_H$, until it becomes almost as large as when it is excited by
the uniform RF field $h_\text{rf} \simeq 2.1$~Oe used in
FIG.~\ref{fig:12}a.

The experimental data and their analysis presented in the previous
sections~\ref{sec:MRFM} to \ref{sec:theory} demonstrate that in the
exact perpendicular configuration, only $\ell=+1$ modes are excited by
the RF current flowing through the nano-pillar, due to the orthoradial
symmetry of the induced RF Oersted field, Eq.~(\ref{eq:irf}). Because
there is no overlap between this particular excitation symmetry and
the uniform azimuthal symmetry of the $\ell=0$ modes, the latter do
not couple to the RF current excitation. The fact that these hidden
modes in the exact perpendicular configuration can be excited by
introducing a small misalignment angle between the applied field and
the normal to the nano-pillar $\unit z$ is a striking result. It means
that the selection rules associated to the RF current excitation
change if the applied field is tilted away from $\unit z$, what we
shall now explain.

Due to the smaller demagnetizing field in the thin magnetic disk than
in the thick one (due to $M_a<M_b$), the equilibrium angle of the thin
layer is smaller than in the thick layer, $\theta_a<\theta_b$, as
obtained from Eq.~(\ref{eq:equil}). For the parameters of our
nano-pillar, $\beta=\theta_b-\theta_a \approx 2 \theta_H$, at $H_{\rm
  ext}\approx10$~kOe and for a small angle $\theta_H$. It means that
if $\theta_H\neq 0$, the magnetization vectors in both layers are
misaligned from each other by an angle $\beta=(\bm{M}_a,\bm{M}_b)$, so
that the cross product $\unit u_a \times\unit u_b$ is finite and lies
in the plane parallel to the layers, say along $\unit x$.  Thus, the
spin transfer excitation $(2\pi\lambda)^{-1}
i_\text{rf}\sin{\beta}\unit x$ associated to the RF current flowing
through the spin-valve nano-pillar \cite{sankey06,chen08}, which is
vanishing in the exact perpendicular configuration where $\beta=0$,
becomes finite if there is a small misalignment angle $\theta_H\neq 0$
(see Eqs.~(\ref{eq:st-fmr}) and (\ref{eq:lambda}) in
appendix~\ref{app:motion}, $(2\pi\lambda)^{-1}$ is the spin transfer
efficiency). Because this so-called ST-FMR excitation has the same
symmetry as an in-plane uniform RF magnetic field, it is expected to
excite SW modes having the $\ell=0$-index symmetry. Still, this
excitation has to compete with the RF Oersted field excitation, which
is independent of $\theta_H$ and is much larger in our configuration
due to the small value of $\beta$ ($<5^\circ$). Therefore the
amplitudes of the $\ell=+1$ modes are much larger than those of the
$\ell=0$ modes in FIG.~\ref{fig:12}b.

It is also clear that the amplitude of the mode $B_{00}$ excited by
the RF current (purple peaks in FIG.~\ref{fig:12}b) grows linearly
with $\theta_H$, as expected from the above discussion. We emphasize
that a quantitative understanding of the amplitude of the peaks
excited by ST-FMR would require to consider the collective nature of
the dynamics in the nano-pillar and the asymmetry of spin transfer in
the thick and thin magnetic layers. Finally, we note that the small
signal observed at $\theta_H=0$ should in principle vanish with the
ST-FMR excitation. This reminiscent signal can be ascribed to a small
misalignment of the applied field with respect to the normal to the
nano-pillar (the precision on the orientation is $0.2^\circ$) or to a
slight asymmetry of the RF current lines through the nano-pillar,
which would induce a small asymmetry of the RF Oersted field, thereof
adding a small in-plane uniform component to the orthoradial magnetic
field.

To summarize, this study enables to derive the selection rules of the
RF current excitation. In the exact perpendicular configuration, the
magnetizations of both layers are aligned, and only $\ell=+1$ modes
can be excited due to the orthoradial symmetry of the current-created
Oersted field ($\ell=0$ modes are hidden). But when a finite angle is
introduced between the magnetizations in each layer by slightly
tilting the applied field away from the normal, $\ell=0$ modes can be
excited by ST-FMR, which has the same symmetry as a uniform RF field
excitation polarized in-plane.

\subsection{Confinement asymmetries \label{sec:asym}}

\begin{figure}
  \includegraphics[width=8.5cm]{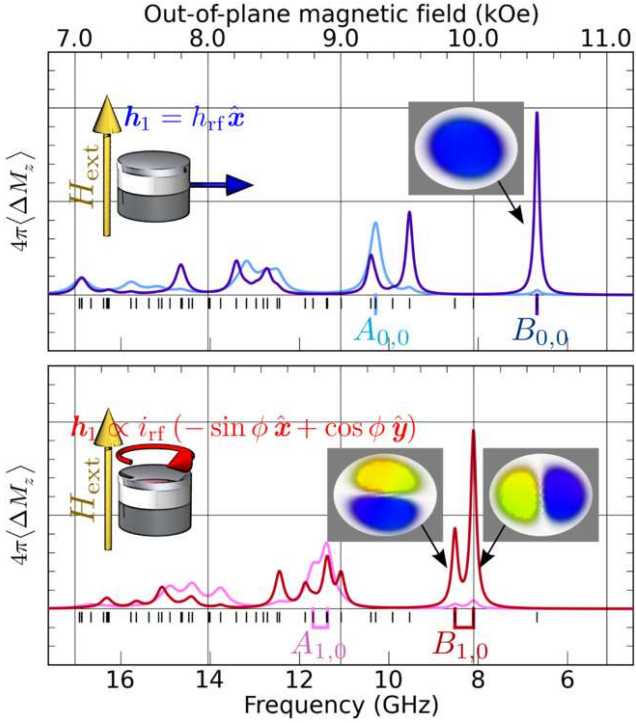}
  \caption{(Color online) Simulated SW spectra for a nano-pillar with
    an elliptical section (see text). Linear response to a homogeneous
    RF magnetic field excitation (a) and to an orthoradial RF Oersted
    field excitation (b). The precession patterns of the lowest energy
    modes are shown in the insets.}
  \label{fig:13}
\end{figure}

As seen in section~\ref{sec:simu}, the 3D micromagnetic simulations
enable to identify the SW modes observed in the experimental
spectra. Still, the latter are more rich than the simulated power
spectra, due to the splitting of some resonance peaks, which was noted
in sections~\ref{sec:hrfvsirf} and \ref{sec:idc}. In particular, the
experimental peak at $H_{\pkA}$, identified as the mode $B_{10}$, is
clearly split in two, with a smaller resonance about 100~Oe away in
the low field wing of the main peak, which is not the case in the
simulation (see FIG.~\ref{fig:10}). The peaks at $H_{\pkB}$
(identified as $B_{11}$) and at $H_{\pkC}$ ($A_{10}$) are also split,
contrary to the simulations, where all these peaks are single. In
contrast to these observations, the peak at $H_{\pka}$, which is the
uniform mode $B_{00}$, is single both in the experimental and
simulated spectra.

So, it seems that experimentally, the occurrence of the mode splitting
depends on the mode index, whereas in 3D simulations, in which the
nano-pillar has a perfect cylindrical shape, none of the resonance
peaks is split. This suggests that the observed splittings are related
to asymmetries in the confinement of the disks, and that the various
SW modes are affected differently because they probe different
regions. The fact that the double peak at $H_{\pkA}$ depends on the
tilt angle (see FIG.~\ref{fig:12}b) and is more or less pronounced
depending on the direction in which the applied field is tilted from
the normal (not shown) is another strong indication that some symmetry
breaking in the lateral confinement is at the origin of this effect
\cite{footnotePigeau}.

To support this idea, we have carried out new 3D micromagnetic
simulations with the SpinFlow 3D package on a structure that break the
perfect cylindrical symmetry of the nano-pillar. We have kept a
perfectly flat structure, but we have used an elliptical
cross-section. The long axis of the ellipse is 250~nm, while the short
axis is 200~nm. The influence of this breaking of symmetry on
respectively the $\ell=0$ spectrum (RF field excitation) and the
$\ell=+1$ spectrum (RF current excitation) is presented in
FIG.~\ref{fig:13}.

Concentrating first on the standard FMR SW spectrum of
FIG.~\ref{fig:13}a, one can see that the lowest energy mode $B_{00}$
remains a single peak. This illustrates the intuitive idea that the
uniform SW mode, where the oscillation power is mostly concentrated at
the center, is rather not sensitive to change of the confinement at
the periphery. The same behavior applies for the lowest energy mode of
the thin layer, $A_{00}$.

The simulated SW spectrum of FIG.~\ref{fig:13}b shows a different
sensitivity to the shape asymmetry. It is observed that the lowest
energy mode with the $\ell=+1$-index splits in two peaks, in contrast
with the lowest energy $\ell=0$ mode which remains a single peak. We
also note that the satellite peak, induced by the elliptical
confinement, is located in the low field wing of the main resonance,
as in the experiments of FIG.~\ref{fig:02}b. The precession patterns
shown in the two insets FIG.~\ref{fig:13}b reveal that the elliptical
shape introduces some mixing between the $\ell=+1$ and $\ell=-1$ SW
modes (the $\ell=-1$ mode corresponds to SWs that are rotating around
the disk in the opposite direction as the Larmor precession). In a
circular disk, these two modes are degenerate, and only the $\ell=+1$
mode couple to the orthoradial Oersted field excitation.  But in the
ellipse, the two eigen-modes split and become mixed, as shown by the
two eigen-vectors displayed in FIG.~\ref{fig:13}b, which correspond to
the linear combinations $J_{1}(\rho) +J_{-1}(\rho)$ and $J_{1}(\rho)
-J_{-1}(\rho)$. The simulated spectrum of FIG.~\ref{fig:13}b
reproduces well the main features of the mechanical-FMR spectrum of
FIG.~\ref{fig:02}b, including the splittings observed for the
$\ell=+1$ SW modes (revealed even better by injecting a dc current
through the nano-pillar, see FIG.~\ref{fig:03}b). Thus, a small
elliptical shape produced during the nanostructuration of the
nano-pillar is most likely responsible for the double peak observed at
$H_{\pkA}$ in FIG.~\ref{fig:02}b.

To summarize, the comparison between 3D simulations and experiments
demonstrate that the observed mode splittings originates from a small
asymmetry in the lateral confinement of the nano-pillar.

\section{Conclusion \label{sec:concl}}

In summary, we used the MRFM technique \cite{klein08} to study the
SW eigen-modes in the prototype of a STNO - a normally magnetized
nano-pillar composed of two magnetic layers coupled by dipolar
interaction.

In contrast to transport spectroscopy techniques
\cite{sankey06,petit07}, MRFM is sensitive to all SW modes excited in
the sample \cite{loubens05} and is completely independent of the
transport properties of the studied spin-valve sample.  Therefore,
MRFM provides an alternative and complementary view on the
magnetization dynamics in hybrid magnetic nano-structures. The
additional advantages of the MRFM technique are its high sensitivity
(in this study, it was able to detect angles of precession as low as
1$^\circ$ in the thin magnetic layer) and its ability to operate on
standard STNO devices buried under contact electrodes without a
specific probe access to the studied sample \cite{pigeau11}.

Using MRFM, we were able to compare the SW spectra of a passive
perpendicularly magnetized STNO-like sample excited by a uniform
in-plane RF magnetic field and by an RF current flowing
perpendicularly through the layers. We found that distinctly different
SW modes (having azimuthal indices $\ell=0$ and $\ell=+1$,
respectively) are excited by the two above mentioned excitation
methods. By studying the influence of a spin polarized dc current on
the observed SW spectra we were able to determine which of the
magnetic layers of the studied nanopillar plays the dominant role in
the magnetization dynamics resulting in the appearance of each
particular SW mode.

We also developed a simple analytic theory allowing to perform a
comprehensive labeling of all the SW eigen-modes of a magnetic
nanopillar in the studied axially symmetric case. This labeling
requires three independent indices: the usual azimuthal and radial
indices $\ell$ and $n$ used for the SW modes of a single magnetic disk
and an additional index referring to the binding or anti-binding ($B$
or $A$) coupling between the two magnetic disks forming a
nanopillar. The obtained experimental and analytic results were also
compared to the results of 3D micromagnetic simulations obtained with
the SpinFlow 3D package \cite{SpinFlow3D}, which confirmed the mode
labeling obtained from the analytic theory.

Thus, we learned that in the axially symmetric case of a
perpendicularly magnetized nano-pillar, the excitations by the RF
field the RF current lead to two mutually orthogonal (and mutually
exclusive) sets of excited SW modes: only the $\ell=0$ modes are
excited by the uniform RF magnetic field, while only the $\ell=+1$
modes are excited by the RF current. Therefore, the $\ell$-index,
related to the azimuthal symmetry of the SW modes, is the
discriminating parameter for the selection rules of the SW mode
excitation.

Moreover, we have demonstrated experimentally and numerically that the
mode selection rules are affected by the breaking of the axial
symmetry of the studied nano-pillar, either by tilting the bias
magnetic field or by making the sample cross-section elliptical. In
particular, if the axial symmetry is broken by tilting the bias
magnetic field, the $\ell=0$ modes can also be excited by an RF
current. This excitation is caused by the ST-FMR mechanism working
when the magnetization vectors in the two magnetic layers of the
nano-pillar are not collinear. Also, the importance of the dynamic
dipolar interaction between the magnetic layers of the nano-pillar
have been clearly demonstrated by our results.

We believe that our results are important for the optimization of the
characteristics of nano-spintronic devices, and in particular STNOs,
and for the experimental determination of the STNO parameters.

First of all, an accurate identification of the SW modes that can be
excited in an STNO nano-pillar is necessary to understand the details
of the high frequency STNO dynamics. The proposed identification of
the nano-pillar SW modes can be used for the experimental
determination of the nano-pillar characteristics, such as radius,
static magnetization, gyromagnetic ratio, and dissipation. We note
that in traditional STNO experiments, where the magnetization dynamics
in a magnetic nano-pillar is excited by a spin-polarized bias current
creating a significant Oersted magnetic field with the $\ell=+1$
symmetry, it is easy to mix-up the $\ell=0$ and the $\ell=+1$ SW
eigen-modes. The spectra of these modes are in almost translational
correspondence, and the experimentally observed dependence of the mode
frequencies on the bias magnetic field can be well described by the
traditional Kittel expression (see \textit{e.g.}
\cite{kiselev03}). Thus, the possibly excited $\ell=+1$ mode can be
easily interpreted as a $\ell=0$ mode, which will lead to the apparent
reduction of the ``free'' layer static magnetization necessary to fit
the Kittel expression for the mode frequency. For instance,
mislabeling the lowest energy mode of the $\ell=+1$ SW spectrum as the
uniform mode ($\ell,n=0,0$) combined with a small misalignment of the
applied field would lead to a discrepancy as large as 1~kG in our
case.

Second, the fact that in most cases both magnetic layers of a
nano-pillar take part in current-induced magnetization dynamics is
very important for the correct identification of the excited SW
modes. The collective (coupled) character of the SW modes in a
nano-pillar can directly influence the magnitude of the spin transfer
torque, which is dependent on the relative orientation of the
magnetization vectors in the two magnetic layers. One might expect,
that the efficiency of the spin transfer torque for a particular SW
mode depends not only on which layer (``free'' or ``fixed'') is
dominating the mode dynamics, but also on the coupling (in phase or
anti-phase) between the magnetization precession in two layers. In our
experimental case, the interlayer coupling is in-phase for the SW
modes dominated by the dynamics in a ``free'' (thin) layer. To obtain
an interlayer coupling which is anti-phase for the SW modes dominated
by the dynamics of the thin layer would require for example to
increase its magnetization compared to the ``fixed'' layer one.

Finally, it is important to note that the MRFM technique has allowed
us to study spin transfer effects in the axially symmetric
configuration of a perpendicularly magnetized nano-pillar, where the
excitation of magnetization dynamics by ST-FMR vanishes due to the
symmetry reasons. This geometry is rather important for applications
as the excited SW modes have the maximum non-linear frequency shift
coefficient \cite{slavin09}. This creates the maximum agility of the
mode frequency with the bias current and, therefore, the maximum width
of the synchronization band to the external periodic signal and to the
large arrays of other STNOs \cite{slavin09}. Phase synchronization has
been identified as a possible mean to dramatically increase the
generated microwave power of these nano-oscillators and, at the same
time, reduce their linewidth \cite{kaka05, mancoff05, slavin05a,
  grollier06, georges08, ruotolo09, urazhdin10, dussaux11}.

\begin{acknowledgments}
  This research was partially supported by the European Grant Master
  (NMP-FP7 212257) and by the French Grant Voice (ANR-09-NANO-006-01),
   by the contract from the U.S. Army TARDEC, RDECOM, and by the grants
ECCS-1001815 and DMR-1015175 from the National Science Foundation of
the USA.
\end{acknowledgments}

\appendix

\section{Theoretical material \label{app:theo}}

\subsection{Equation of motion \label{app:motion}}

In this appendix, we detail the derivation of the equations which
govern the dynamics of a ferromagnetic layer in the presence of an
external periodic excitation and of spin transfer, following the
general formalism introduced in section~\ref{sec:general}. For an
isolated layer, the \emph{local} dynamics (within the exchange length)
of the magnetization vector is described by the Landau-Lifshitz (LL)
equation:
\begin{equation}\label{LL-pert}
  \frac{1}{\gamma}\frac{\partial\bm M}{\partial t} =
  \bm H \times \bm M + \bm h(t) \times \bm M \, , 
\end{equation}
with $\gamma$ being the modulus of the gyromagnetic ratio. The LL
equation is written here in its perturbative form, where the second
term on the right-hand-side of Eq.~(\ref{LL-pert}) represents the
perturbation term. The field $\bm H$ is the effective magnetic field:
\begin{equation}\label{WB-a}
  \bm H  = \bm H_0 - 4 \pi \op G * \bm M 
  \ .\end{equation}
Here, $\bm H_0$ is the total static external magnetic field (possibly
spatially-dependent) and the linear tensor self-adjoint operator $\op
G$ describes the magnetic self-interactions.

Considering only the linear processes, we can represent the time
dependent (out-of-equilibrium) part of the magnetization as a series
over the SW eigen-modes:
\begin{eqnarray}\label{M-series}
  \bm M(t, \bm r) - M_s \unit u(\bm r) &\approx& \bm m (t, \bm r)
  \\\nonumber
  &\approx& \sum_\nu c_\nu(t)\bm m_\nu(\bm r)
  + {\rm c.c.} \, ,
\end{eqnarray}
where $M_s$ is the saturation magnetization of the layer. Here, ${\rm
  c.c.}$ stands for the complex-conjugated part. The coefficients
$c_\nu(t)$ are time-dependent SW amplitudes.

The second term on the left-hand-side of Eq.~(\ref{LL-pert})
represents the perturbations from the equilibrium state, including the
non-adiabatic contributions. The non-conservative perturbation
magnetic field $\bm h (t)$ may depend on time and be a function of the
magnetization distribution $\bm M(t)$. It can be approximately
represented as:
\begin{equation}\label{b-expansion}
  \bm h (t) = \bm h_1 (t) + \op L_1 * \bm m (t)
  \,,
\end{equation}
where $\bm h_1 (t)$ is the external perturbation field and $\op L_1$
is a certain linear operator, allowing $\bm h (t)$ to depend on the
magnetization distribution. The latter case may describe the influence
of the Gilbert damping $-(\alpha/\gamma M_s)\partial\bm M/\partial t$
through:
\begin{equation}
  \op L_1 * \bm m = i\frac{\alpha}{\gamma}\sum_\nu \omega_\nu(c_\nu(t)\bm m_\nu(\bm r) - {\rm c.c.})
  \,,\end{equation}
where $\alpha$ is the Gilbert damping constant.

Substituting the series representation Eq.~(\ref{M-series}) and the
representation of the perturbation field Eq.~(\ref{b-expansion}) into
Eq.~(\ref{LL-pert}) and using orthogonality relations
Eq.~(\ref{ortho}), one can obtain the following equations for the SW
amplitudes $c_\nu$:
\begin{equation}\label{eq:dcdt-1}
	\frac{dc_\nu}{dt} = -i\omega_\nu c_\nu +i \gamma\sum_{\nu'}\left(S_{\nu,\nu'}c_{\nu'} + S_{\nu,\overline{\nu'}}\overline{c}_{\nu'}\right) + i \gamma h_\nu
\,,\end{equation}
where
\begin{subequations}
\begin{eqnarray}
  S_{\nu,\nu'} &=& \frac{\< \cc m_\nu \cdot \op L_1 * \bm
    m_{\nu'} \> - \<(\unit u \cdot \bm h_1 )(\cc m_\nu \cdot \bm m_{\nu'})\>}{\mathcal{N}_\nu}
  \,,\nonumber\\\\
  S_{\nu,\overline{\nu'}} &=& \frac{\< \cc m_\nu \cdot \op L_1 * \cc m_{\nu'} \> - \<(\unit u \cdot \bm h_1)(\cc m_\nu \cdot \cc m_{\nu'})\>}{\mathcal{N}_\nu}
  \,,\nonumber\\\\
  h_\nu &=& \frac{\<\cc m_\nu \cdot \bm h_1\>}{\mathcal{N}_\nu}
  \ .\end{eqnarray}
\end{subequations}
In many cases the perturbed equations (\ref{eq:dcdt-1}) can be further
simplified by retaining only the diagonal term $S_{\nu,\nu} =
i\alpha\omega_\nu\<\cc m_\nu\cdot\bm m_\nu\>/(\gamma \mathcal{N}_\nu)$
(assuming that there are no degenerate modes). The SW damping rate is
then given by
\begin{equation} \label{eq:gamma} \Gamma_\nu = \alpha \omega_\nu
  \frac{\<\cc m_\nu \cdot \bm m_\nu\>}{\mathcal{N}_\nu} \ .
\end{equation}
The damping rate $\Gamma_\nu$ is responsible for the finite linewidth
of the resonance peaks, $\Delta H$ (FWHM). If the sample is
homogeneously magnetized and the precession is circular, the simple
relation $\alpha \Delta H = \omega_\nu/\gamma$ holds.

From the equations above, one recovers for the coefficient $c_\nu$ the
equation of motion of a damped harmonic oscillator: 
\begin{equation} \label{eq:dynam} 
\frac{dc_\nu}{dt} = -i\omega_\nu c_\nu - \Gamma_\nu
c_\nu + i \gamma h_\nu \,.
\end{equation} 

If a second magnetic layer $j^\prime$ is electrically connected to
layer $j$ and spin transfer is allowed between them, the equation of
motion must be modified.  When a charge current $I$ is flowing through
the layers, the additional Slonczewski-Berger term
\cite{slonczewski96,berger96} adds to the operator $\op L_1$ of
Eq.~(\ref{b-expansion}). It is possible to define an effective damping
in layer $j$, which depends on the spin polarized current and on the
relative angle between the magnetization in layer $j$ and the
direction of the spin polarization $\unit u_{j^\prime}$
\cite{slavin09}:
\begin{equation}
  \Gamma_\nu^{*} = \alpha\omega_\nu \frac{\<\cc m_\nu \cdot \bm
    m_\nu\>}{\mathcal{N}_\nu} + \frac{I \epsilon}{2 e N_s} \frac{
    \<(\cc m_\nu \cdot\bm m_\nu)  (\unit u_j \cdot\unit u_{j^\prime})
    \>}{\mathcal{N}_\nu} \,,
  \label{eq:stt}
\end{equation}
where
\begin{equation}
  N_s=\frac{M_j V_j}{\gamma \hbar}
  \label{eq:Ns}
\end{equation}
is the dimensionless total number of magnons that can be excited
inside the volume $V_j$ of layer $j$. Here, $\hbar$ is the reduced
Planck constant, $e$ the modulus of the electron charge and $\epsilon$
the spin polarization efficiency of the current. The threshold current
for auto-oscillations in layer $j$ corresponds to $\Gamma_\nu^{*}=0$,
\textit{i.e.}, if $\unit u_j \parallel \unit u_j^\prime$,
$I_\text{th}=-2 \alpha\omega_\nu N_s e/ \epsilon$. Using
$\epsilon=0.3$ and the parameters of our thin layer, one can estimate
$I_\text{th}\simeq -4.8$~mA for the uniform SW mode at $8.1$~GHz, in
agreement with the experimental data \cite{footnoteHamadeh}. This
result is also in quantitative agreement with calculations performed
for our nano-pillar device with no adjusting parameters in the
framework of continuous random matrix theory (CRMT) described in
Ref.\cite{rychkov09}.

We now turn to the periodic external excitation $\bm{h}_1(t) =
\bm{h}_1 \exp^{i \omega t}$, whose amplitude $\bm{h}_1$ is composed of
three different contributions,
\begin{equation}
  \bm{h}_1 = 
  \bm{h}_\text{u}+\bm{h}_\text{Oe}+\bm{h}_\text{ST} \,,
  \label{eq:driving}
\end{equation}
that we shall detail below.

The first type of excitation corresponds to a uniform RF magnetic
field applied perpendicularly to the effective field $\bm{H}$. This
configuration corresponds to conventional FMR spectroscopy. Assuming
that the sample is uniformly magnetized along the nano-pillar symmetry
axis $\unit z$, it reduces to:
\begin{equation}
  \bm{h}_\text{u}={h}_\text{rf}\unit x,
  \label{eq:hrf}
\end{equation}
where $\unit x$ is a unit vector in the in-plane direction and
${h}_\text{rf}$ the linearly polarized amplitude.

SW spectroscopy can also be performed by injecting a uniform RF charge
current $i_\text{rf}$ through the nano-pillar (\textit{i.e.}, along
$\unit z$). First, this produces an orthoradial RF Oersted
field:
\begin{equation}
  \bm{h}_\text{Oe}=\left[\frac{4 \pi}{10}\right]
  \frac{i_\text{rf}}{2 \pi R} \frac{\rho}{R}(-\sin\phi\,\unit x +
  \cos\phi\,\unit y) \ ,
  \label{eq:irf}
\end{equation}
where $R$ is the radius of the nano-pillar and $(\rho,\phi)$ are the
polar coordinates. In this formula, the current should be expressed in
A and the prefactor between the square brackets converts A/cm into Oe
(cgs units). The maximum amplitude of the RF Oersted field is reached
at the periphery of the nano-pillar, $\rho=R$, and equals 1.6~Oe for a
peak amplitude $i_\text{rf}=100$~$\mu$A and the experimental
parameters.

Second, the RF current produces a ST-FMR excitation:
\begin{equation}
  \bm{h}_\text{ST}= \frac{i_\text{rf}}{2 \pi\lambda} \left[\unit u_{_j}
    \times \unit u_{j^\prime} \right]. 
  \label{eq:st-fmr}
\end{equation}
where we have rewritten the spin-transfer efficiency of the charge
current in Eq.~(\ref{eq:stt}) as a function of
\begin{equation}
  2\pi \lambda= \gamma \frac{2 e N_s}{\epsilon},
  \label{eq:lambda}
\end{equation}
which has the dimension of a distance ($\lambda\simeq 200$~nm for our
thin layer).  If the thin and thick layers are misaligned by an angle
$\beta$ in the plane ($y$,$z$), Eq.~(\ref{eq:st-fmr}) reduces to
$\bm{h}_\text{ST}=i_\text{rf}/(2\pi\lambda)\sin{\beta} \unit x$, which
demonstrates that the ST-FMR excitation is equivalent to a linearly
polarized RF magnetic field, Eq.~(\ref{eq:hrf}). The ST-FMR excitation
vanishes if the magnetic layers are parallel. The amplitude ratio
between the ST-FMR and the RF Oersted field excitations, both produced
by the RF current flowing through the nano-pillar, is
$h_\text{ST}/h_\text{Oe}\simeq(R/\lambda)\sin{\beta}$. In our
geometry, $\lambda \approx 2\,R$, but due to the small angle $\beta$
between the layers, the RF Oersted field contribution is much larger
than the ST-FMR one. We note that even if $\beta\approx\pi/2$, the
contribution of the RF Oersted field cannot be disregarded in general
in ST-FMR experiments.

\subsection{Numerical application \label{app:na}}

In this section, we derive a practical guideline to calculate the
eigen-frequencies $\omega$ using the analytical formalism developed in
section~\ref{sec:analytical}. Let ${\bm m}_{\nu}$ be a certain
orthogonal basis in the space of the vector functions ${\bm m}$
satisfying both the local orthogonality to $\unit u$ and the total
pinning condition at the boundary of the magnetic body. Spin-wave
eigen modes can thus be expressed as a series expansion on the ${\bm
  m}_{\nu}$ basis (cf. Eq.~(\ref{M-series})). A general expression for
the eigen-frequencies can be found from the condition of vanishing
determinant:
\begin{equation}\label{omega-det}
  \Big|\Big|\, \mathcal{N}_{\nu^\prime,\nu} \omega -
  \mathcal{N}_{\nu^\prime,\nu} \left\{ \op \Omega \right\}_
  {\nu^\prime,\nu}  \,\Big|\Big| = 0 \ ,
\end{equation}
where for the simplicity of the discussion, it is convenient to
introduce a curly bracket notation, to indicate that the enclosed
quantity is spatially weighted by the spatial pattern of the mode
profile and averaged:
\begin{equation}\label{eq:omega-nu}
  \left\{ \op \Omega \right\}_{\nu^\prime,\nu} \equiv \frac{\<{\cc
      m}_{\nu^\prime} \cdot \op\Omega * {\bm m}_{\nu}\>}{\mathcal{N}_{\nu^\prime,\nu}} \,,
\end{equation}
This echoes the chevron bracket notation introduced in
Eq.~(\ref{eq:average}) to indicate the homogeneous spatial average
over the volume of the magnetic body. Here $\mathcal{N}$ represents a
renormalization quantity, defined by
\begin{equation}\label{NLjk-a}
\mathcal{N}_{\nu^\prime,\nu} \equiv i\<{\cc m}_{\nu^\prime} \cdot (
\unit u \times {\bm m}_{\nu})\> \ ,
\end{equation}
which has in general off-diagonal elements.

In the case of perpendicularly magnetized disks, where the set of
Bessel functions $\frac 12 (\unit x + i \unit y)e^{-i\ell\phi}
J_\ell(k_{\ell,n} \rho)$ diagonalizes the uncoupled Hamiltonian, the
secular Eq.~(\ref{omega-det}) becomes diagonal and we recover
Eq.~(\ref{omega-integral}):
\begin{equation}\label{eq:curly}
  \omega_\nu=\left\{ \op \Omega \right\}_{\nu,\nu} 
\end{equation}
 
We shall now perform the numerical application of the eigen-value of
the lowest energy mode ($\ell,n=0,0$) using the parameters of our
nano-pillar shown in Table~\ref{tab:param}. We will drop the subscript
$\nu$ to the curly brackets, understanding that the spatial average in
Eq.~(\ref{eq:curly}) is made over the uniform mode $\bm m_\nu = \frac
12 J_0(k_0 \rho) (\unit x + i \unit y)$, where $k_0=2.4048/R$ is its
wave-vector. In this case the value of the normalization constant is
simply $\mathcal{N}_0 = \<J_0^2\> = J_1^2(k_0 R) = 0.2695$.

The different contributions that enter inside the operator $\left\{
  \op \Omega\right\}$ are detailed in Eq.~(\ref{eq:op-L}):
\begin{equation}\label{eq:Curly}
  \left\{ \op \Omega \right\} = \gamma \left\{  H
  \right\} + 4 \pi \gamma M_j \left\{ \op G
  \right\} \,.
\end{equation}
We start with the calculation of the amplitude of effective magnetic
field, the first term on the right hand side of
Eq.~(\ref{eq:Curly}). As shown by Eq.~(\ref{eq:heff}), the scalar
value $H$ along $\unit z$ decomposes itself in two terms:
\begin{equation}\label{eq:Heff} 
  \left\{ H \right\} = \left\{ \unit z \cdot \bm H_0
  \right\} - 4 \pi M_s \left\{ \unit z \cdot \op G \ast \unit z\
  \right\}.
\end{equation}
The term $\unit z \cdot \op G \ast \unit z$ represents the static
magnetic self-interaction. In the case of homogeneously magnetized
body, the inhomogeneous exchange contribution to the static
self-interaction is strictly null and the second term of
Eq.~(\ref{eq:Heff}) reduces to the magneto-dipolar contribution $\op
G^{(d)}$, which has the following form in the wave-vector
representation:
\begin{equation}\label{N}
  \op G^{(d)}(\bm r) = \int D(\bm k) \frac{\bm k \otimes \bm
    k}{k^2} \exp^{i \bm k \cdot \bm r} d^3\bm k\, ,
\end{equation}
where $ D(\bm k)$ is the Fourier transform of the body shape function
\cite{beleggia03} and the symbol $\otimes$ denotes direct product of
vectors. For a disk of radius $R$ and thickness $t$, an analytical
expression for the different position-dependent demagnetization tensor
elements of a disk $N_{{uv}\,[R,t]}(\bm r) \equiv \unit u (\bm r)
\cdot \op G^{(d)} \ast \unit v$ valid in the whole space are available
in Ref.\cite{tandon04}. For perpendicularly magnetized disks where
$\unit u = \unit v =\unit z$, the expression of the self-integral
becomes
\begin{equation} \label{eq:self} 
  \left\{ N_{zz}^{(j,j)} \right\} = \frac{1}{\<J_0^2\>} \int_{V_j} \dv J_0^2(k_0 \rho) N_{{zz}\,[R,t_j]} (\rho,z) \, ,
\end{equation}
for both the thin ($j=a$) and thick ($j=b$) layers. Their numerical
values are displayed in Table. \ref{tab:na}.

The term $\left\{ \unit z \cdot \bm H_0 \right\}$ of
Eq.~(\ref{eq:Heff}) is the projection on the precession axis of the
total applied magnetic field. It comprises the external magnetic field
$\left\{ H_\text{ext} \right\} = H_\text{ext}$, the stray field of the
mechanical-FMR probe $\left\{ H_\text{sph} \right\} =190$~Oe and the
cross-magneto-dipolar static interactions between each layer. The
latter can be estimated from the cross tensor elements of the static
magneto-dipolar field of the $j^\prime$-th disk produced over the
volume of the $j$-th disks:
\begin{equation} \label{eq:cross} \left\{ N_{zz}^{(j,j^\prime)}
  \right\} = \frac{1}{\<J_0^2\>} \int_{V_j} \dv
  J_0^2(k_0 \rho) N_{{zz}\,[R,t_{j^\prime}]}(\rho,z+z_0)\, ,
\end{equation}
where $z_0$ is the distance between the centers of the two axially
aligned disks. The numerical values of the cross tensor elements are
reported in Table. \ref{tab:na}.  Putting all the above elements
together, the total effective field simply writes:
\begin{equation}\label{eq:hkittel}
  \left\{ H \right\}  =  H_\text{ext} + \left\{H_{\rm sph}\right\} - 4 \pi \left\{
    N_{zz}^{(j,j)} \right\} M_j- 4 \pi \left\{ N_{zz}^{(j,j^\prime)}
  \right\} M_{j^\prime} \,.
\end{equation}

We now turn our attention to the integration $4 \pi M_j \left\{ \op G
\right\}$, the second term on the right hand side of
Eq.~(\ref{eq:Curly}). We recall that for Py the operator $\op G = \op
G^{(e)} + \op G^{(d)}$ is the sum of the inhomogeneous exchange and
magneto-dipolar interactions.  In the wave-vector representation, $\op
G^{(e)} = \Lambda_{\rm ex}^2k^2\op I$, where the exchange length
$\Lambda_{\rm ex}=\sqrt{2 J/(4\pi M_{j}^2)}$ depends on the exchange
stiffness constant $J$, expressed in erg/cm ($=10^{-6}$ in Py). It
produces the exchange field:
\begin{equation}
  \left\{ H_\text{ex}\right\} =  4 \pi M_j \Lambda_{\rm ex}^2 k_{0}^2 \ ,
\end{equation}
which yields the value $\left\{ H_\text{ex}\right\} = 110$~Oe.

The other contribution is the dynamic magneto-dipolar
self-interaction, which represents the depolarization field of the SW
mode on itself. For the $\ell=0$ modes, an analytical expression can
be derived:
\begin{equation}
  \left\{ N_{xx}^{(j,j^\prime)} \right\}   = 
  \frac{1}{\<J_0^2\>} \int_{V_j} \dv J_0(k_0 \rho) \int_0^{R} \!\!\!\!du
  \frac{\partial N_{{xx}\,[u,t_{j^\prime}]}(\rho,z)}{\partial u} J_0(k_0 u)\, , \label{eq:overlap}
\end{equation}
where the quantity in the second integral is the magnetic stray field
produced at the spatial position $\bm{r}$ by a cylindrical tube of
width $du$, radius $u$, and thickness $t_{j^\prime}$, homogeneously
magnetized \cite{beleggia04} along $\unit x$ by $J_0(k_0 u)$. We use
the same expression above to write the self- and cross-contribution,
understanding implicitly that the spacer value $z_0$ should be added
in the later case, as shown in Eq.~(\ref{eq:cross}). The values of the
self- and cross-tensor elements are reported in the last line of
Table. \ref{tab:na}. We mention, that an approximate expression of the
self-Eq.~(\ref{eq:overlap}) has been derived by Kalinikos and Slavin
\cite{kalinikos86} for the lowest SW branch of platelet shape bodies
with uniform magnetization across the film thickness. This expression
reduces to
\begin{equation}
  \left\{ N_{xx}\right\}  \simeq  \frac{1}{2} \left(1 -G_{0}^\perp\right) \,
\end{equation}
where the analytical expression of $G_{\ell,n}^\perp$ for Bessel
functions is given by Eq.~(26) in ref\cite{klein08}.

The cross elements are responsible for the dynamic dipolar coupling
detailed in section~\ref{sec:dipolar}.  
\begin{equation}
h_{j,j^\prime}=4 \pi \left\{ N_{xx}^{(j,j^\prime)} \right\} M_{j^\prime}
\end{equation}
The value of the coupling frequency $\Omega$ for the lowest energy
mode yields:
\begin{equation}
  \Omega \simeq \gamma \sqrt{\left\{  N_{xx}^{(a,b)} \right\}  4\pi M_b \left\{ N_{xx}^{(b,a)} \right\}  4\pi M_a  }\,,
\end{equation}
which leads to $\Omega/2\pi \simeq 0.56~{\rm GHz}$.

Neglecting the dynamical dipolar coupling (the generalization to
$\Omega\neq 0$ is Eq.~(\ref{eq:AB})), we derive an expression for the
eigen-value of index $\nu = j_{0,0}$:
\begin{equation} \label{eq:omega} \frac{\omega_\nu}{\gamma} = \left\{ H
  \right\} + 4\pi \left\{ N_{xx}^{(j,j)} \right\} M_j +\left\{ H_\text{ex}\right\} \,
  ,
\end{equation}
$j=a,b$ being the layer index and $ \left\{ H \right \} $ being
defined in Eq.~(\ref{eq:hkittel}).  Eq.~(\ref{eq:omega}) is a
simplified expression valid for circularly polarized modes ($\ell=0$
index), where we have taken advantage of the equality $ \{ N_{xx} \}_0
= \{ N_{yy}\}_0$ in our circular disk. This expression can be extended
to higher order modes by using $ \{ N_{xx} \}_{\ell,n} \approx \{
N_{xx} \}_0 \sqrt{k_{\ell,n}/k_0}$ inside Eq.~(\ref{eq:omega}). This
approximation is derived from the ellipticity of $\ell \neq 0$ modes
($ \{ N_{xx} \} \neq \{ N_{yy}\}$). One needs thus two separate
equations (\ref{eq:omega}) for the values of $\omega$ for each
cartesian axis \cite{klein08}: one proportional to $m_x^2/(m_x m_y)$,
the other to $m_y^2/(m_x m_y)$. The product of these two equations is
independent of the ellipticity, leading to the general expression for
the eigen-value of arbitrary index $\nu=j_{\ell,n}$:
\begin{eqnarray}\label{eq:omegaK}
    \frac{\omega_\nu^2}{\gamma^2} & = & \left
      ( \left\{ H \right\}_\nu + 4\pi \left\{ N_{xx}^{(j,j)} \right\}_\nu M_j +
      \left\{ H_\text{ex}\right\}_\nu \right) \times \nonumber \\
    && \left ( \left\{ H \right\}_\nu +
      4\pi \left\{ N_{yy}^{(j,j)} \right\}_\nu M_j+ \left\{
        H_\text{ex}\right\}_\nu \right) \,,
\end{eqnarray}
which can be seen as a generalization of the Kittel formula for
arbitrary shaped multi-body.

Equating $H_{a_{00}}=H_{\pkc}$ and $H_{b_{00}}=H_{\pka}$ in
Eq.~(\ref{eq:omega}), where $H_{a_{00}}$ and $H_{b_{00}}$ are the
resonance fields at $f_\text{fix}=8.1$~GHz of the uniform modes in the
thin and thick disks, respectively, leads to $4\pi M_a=8.0 \times
10^3$~G and $4 \pi M_b=9.6 \times 10^3$~G.

\begin{table}
  \caption{Values of the self- and cross- depolarization tensor
    elements weighted by the precession profile of the uniform mode
    for the thin ($j=a$) and thick ($j=b$) disks.}
  \begin{ruledtabular}
    \begin{tabular}{c c c c c}
       & $(a,a)$   & $(a,b)$ & $(b,a)$ & $(b,b)$ \\
$\{ N_{zz}^{(j,j^\prime)} \}$ & $+0.979$ & $-0.068$ & $-0.017$ & $+0.919$ \\
$\{ N_{xx}^{(j,j^\prime)} \}$ & $+0.016$ & $+0.042$ & $+0.011$ & $+0.056$ \\
    \end{tabular}
  \end{ruledtabular}\label{tab:na}
\end{table}

Finally, the above formalism also enables to determine the angle
$\theta_j=(\unit z,\unit u_j)$ between the equilibrium direction of
the magnetization in layer $j$ and the normal axis when the bias field
is applied at a polar angle $\theta_H=(\unit z,\bm{H}_{\rm
  ext})$. From the equality $ \unit u \times \bm H_\text{ext} = 4\pi
M_j \unit u \times \<\op G^{(d_j)}\ast \unit u \> $, one extracts the
relationship:
\begin{equation}
  H_{\rm ext}\sin{(\theta_j-\theta_H)}=2\pi M_j\left(\langle{N_{zz}^{(jj)}}\rangle
    -\langle{N_{xx}^{(jj)}}\rangle\right)\sin{2\theta_j}
  \label{eq:equil}
\end{equation}
The angle $\theta_j$ is useful to estimate the shift to lower field of
the FMR spectrum \cite{klein08}, $2\pi
M_j(\{N_{zz}^{(j)}\}-\{N_{xx}^{(j)}\}) (1-\cos{2 \theta_j})\approx
420$~Oe, when $\theta_b = 13^\circ$ in the thick layer.

\section{Methods and calibration  \label{app:methods}}

\subsection{Mechanical vibration amplitude \label{app:Mz}}

Here, we detail the experimental protocol used to calibrate the
amplitudes of the uniform RF magnetic field and of the mechanical-FMR
signal. The procedure uses the non-linear properties of the
magnetization dynamics and consists in studying the power dependence
of the line shape. In the following, $\Delta H$ denotes the FWHM
linewidth measured in the linear regime.

We use the onset of foldover as a mean to calibrate the strength of
the RF field produced by the microwave antenna. This non-linear effect
is responsible for the asymmetric shape of the resonance peaks in
FIGS.~\ref{fig:02} and \ref{fig:03}. In fact, it was pointed out by
Anderson and Suhl \cite{anderson55} that the resonance curve at high
power should be skewed, due to the static change of the magnetization
$M_z$, which also shifts the resonance frequency. For a normally
magnetized sample, this non-linear frequency shift is positive
(blue-shift), and the field-sweep line shapes are distorted towards
low field. There is a critical strength of the RF magnetic field $h_c$
(linearly polarized amplitude) for which the slope of the resonance
curve becomes infinite on the low field side of the resonance
\cite{schlomann59a}:
\begin{equation}
  h_c=  2 \Delta H \sqrt{\frac{2 \Delta H}{3 \sqrt{3}
      \left| \left\{ N_{zz} \right\} - \left\{ N_{xx}
        \right\} \right| 4 \pi M_s}} \label{eq:foldover}
\end{equation}
where $ \left\{ N_{zz} \right\} - \left\{ N_{xx} \right\} $ is the
difference between the depolarization factors in the longitudinal and
transverse directions. Experimentally we find that for the peak at
$H_{\pka}$, this onset is reached when the output power of the
synthesizer at $8.1$~GHz is $P_0= +9$~dBm. Using the magnetic
properties of the thick layer (Table~\ref{tab:param}), we infer from
Eq.~(\ref{eq:foldover}) that at the critical onset of foldover, the
strength of the RF magnetic field is $h_c=4.2\pm0.8$~Oe. We note, that
this value is in agreement with the estimation made by directly
evaluating the field produced by the RF current flowing in the antenna
at $8.1$~GHz for this output power, $h_\text{rf}=5.5\pm1$~Oe.

Furthermore, this procedure gives a calibration of the amplitude of
the mechanical-FMR signal $\langle \Delta M_z\rangle$. At the onset of
foldover, the longitudinal change of the magnetization is indeed
\cite{anderson55}
\begin{equation}
  \label{eq:deltamz}
  4 \pi \langle \Delta M_z\rangle=\frac{4}{3 \sqrt{3}} \Delta H.
\end{equation}
A numerical application of Eq.~(\ref{eq:deltamz}) yields $4 \pi
\langle \Delta M_z\rangle=36\pm4$~G, which corresponds to the critical
angle of precession $\langle \theta_c \rangle=5^\circ$. We have used
this calibration of the cantilever vibration amplitude to evaluate the
change of the longitudinal magnetization at the maximum of the peak at
$H_{\pka}$ in FIG.~\ref{fig:02}a.

\subsection{Microwave setup  \label{app:RF}}

In this appendix, we give some details on the microwave circuit, which
was carefully design to minimize the cross-talk between the RF field
and RF current excitation parts. 

The calibration of the RF magnetic field produced by the microwave
antenna has been presented in the previous appendix. In order to
calibrate the RF current flowing through the nano-pillar with respect
to the synthesizer output power injected into the contact electrodes,
we have first used a standard microwave setup. The nano-pillar
electrodes are directly connected to the microwave synthesizer through
a picoprobe, a bias-T, and a semi-rigid coaxial line, that allow to
perform voltage-FMR spectroscopy. In this experiment, the amplitude of
$i_\text{rf}$ flowing through the nano-pillar can be accurately
determined, owing to the determination of losses and reflexions in the
microwave circuit using a network analyzer. Then, the same experiment
is repeated inside the MRFM setup, in which the contact electrodes are
wire bounded to a microwave cable and the circuit contains more
connections. The comparison with the standard setup yields an
estimation of the rms amplitude of the RF current in the
mechanical-FMR setup: $i_\text{rf}=170 \pm40$~$\mu$A for an output
power of $-22$~dBm injected at $8.1$~GHz through the contact
electrodes.

It is also possible to estimate experimentally the high frequency
coupling between the microwave antenna and the electrodes that contact
the nano-pillar. For this, we exploit the fact that in the exact
perpendicular configuration, different SW modes are excited by the
uniform RF field ($\ell=0$-index) and by the RF current
($\ell=+1$-index). If the RF magnetic field used to excite the
conventional FMR spectrum would induce any relevant RF eddy current
through the nano-pillar, $\ell=+1$ modes which are excited in the SW
spectrum of FIG.~\ref{fig:02}b should also be detected in the SW
spectrum of FIG.~\ref{fig:02}a, which is not the case. We deduce from
this observation that for an output power of $+3$~dBm injected in the
antenna, the induced eddy current through the nano-pillar is less than
when injecting $-38$~dBm directly through the contact electrodes,
\textit{i.e.}, $i_\text{rf}< 30~\mu$A. So, at
$f_\text{fix}=8.1$~GHz, the isolation between the two parts of the
microwave circuit is better than 40~dB. However, we note that the
latter depends on the frequency, and that for some particular values,
it can drop to only 20~dB.

Still, owing to the broadband design of the contact electrodes and to
the low microwave power required to excite SW modes with the RF
current excitation part, it is possible to acquire FMR spectra at a
fixed bias magnetic field $H_\text{fix}$ by sweeping the frequency of
the RF current. We mention that in the frequency-sweep experiments
presented in FIG.~\ref{fig:04}a, the output power of the synthesizer
is kept at $-22$~dBm over the full frequency range (4 to 18~GHz),
which results in an amplitude variation of $i_\text{rf}$, mainly
associated to frequency dependent losses in the circuit. We also note
that the same frequency-sweep experiment cannot be performed as
cleanly with the RF field excitation due to the high power that has to
be injected in the microwave antenna and to the dependence of the
isolation on frequency mentioned above.

\subsection{Cavity-FMR characterization of the extended
  film \label{app:cavity}}

Before the nano-fabrication of the nano-pillar devices, a reference
film of Cu60 $|$ Py$_b$15 $|$ Cu10 $|$ Py$_a$4 $|$ Au25 (thicknesses
in nm) is cut out from the Si wafer for characterization purpose. The
extraction of the material parameters is obtained independently on
this reference film by a reflexion X-band spectrometer (9.6~GHz)
operating at room temperature. The experiment consists in measuring
the resonance spectra of the multi-layer as a function of the polar
angle $\theta_H$ between the applied field and the normal to the
film. The resonance field of the layer $j=a,b$ as a function of
$\theta_H$ depends only on the gyromagnetic ratio $\gamma$ and on the
total perpendicular anisotropy field, which here reflects entirely the
demagnetizing field $4 \pi M_j$ of the layer \cite{hurdequint02}. The
obtained values for the gyromagnetic ratio (identical for both layers)
and the magnetizations are collected in Table~\ref{tab:param}. The
magnetization of the thick layer (9.6~kOe) corresponds to the expected
value for bulk Py with composition Ni$_{80}$Fe$_{20}$. The
magnetization of the thin layer is 1.4~kG smaller, which reflects the
reduction of the magnetization in the interfacial layer (of the order
of 1~nm), due to the gradual composition variation of the NiFe alloy
from Ni$_{80}$Fe$_{20}$ to the normal metal (Cu or Au)
\cite{mizukami01,hurdequint07}.

An estimate of the damping parameter and the amount of inhomogeneous
broadening can also be obtained from the angular dependence of the
linewidth of the resonant mode associated to each layer. The linewidth
is in general the sum of two contributions: an intrinsic relaxation of
the magnetization vector (homogeneous width) and an inhomogeneous
broadening corresponding to a distribution of resonance fields (whose
main sources have been described for a Permalloy polycrystalline layer
\cite{hurdequint02}). The intrinsic damping parameter is deduced from
the parallel geometry linewidth. In the Py$_b$ 15~nm thick layer, the
linewidths observed in the parallel and perpendicular geometries are
respectively: $\Delta H_{\parallel b}=64$~Oe and $\Delta H_{\perp
  b}=73$~Oe. The higher value observed in the perpendicular geometry
reveals a fair amount of inhomogeneities \cite{hurdequint02}. The
linewidth observed in the parallel geometry corresponds to an
intrinsic damping parameter $\alpha_{b}=(0.9\pm0.1) \times
10^{-2}$. For the thin Py$_a$ 4~nm layer, $\Delta H_{\parallel
  a}=83$~Oe and $\Delta H_{\perp a}=171$~Oe. From the parallel
geometry linewidth, we deduce an intrinsic damping parameter
$\alpha_{a}=(1.5\pm0.3) \times 10^{-2}$. This value is higher than for
the thick layer because of the larger effect of the diffusion of the
microwave magnetization of the conduction electrons in the adjacent
normal metal layers \cite{hurdequint07}, associated to the fact that
the thin layer thickness (4~nm) is less than the spin-diffusion length
in Py.  The much larger value $\Delta H_{\perp a}$ is associated to a
large contribution of the inhomogeneous broadening arising from a
substantial effect of interfacial roughness (and dispersion of heights
of the crystallites of the base) on the thin Py layer grown on top of
a 85~nm thick metallic base.


\begin{thebibliography}{10}
\bibitem{footnoteGMR}%
  \BibitemOpen
  \bibinfo {note} {We mention here that the measured amplitude of the GMR
  effect in our sample is $\Delta R_\text{GMR}=25$~m$\Omega$, in agreement with
  the value calculated by CRMT for the multi-layer stack composition
  \cite{rychkov09}.}%
  \bibAnnoteFile{Stop}{footnoteGMR}%
\bibitem{footnotemirror}%
  \BibitemOpen
  \bibinfo {note} {The two images of FIGS.~\ref{fig:11}b and \ref{fig:11}c are
  top and bottom view of the nano-pillar, hence one of them should be mirrored
  for direct comparison of the relative phase between the layers.}%
  \bibAnnoteFile{Stop}{footnotemirror}%
\end{thebibliography}
\end{document}